\documentclass[11pt]{article}
\usepackage{latexsym}
\usepackage{amssymb}
\usepackage[dvips]{graphicx}
\usepackage{epsfig}
\usepackage{rotating}
\usepackage{amsfonts}
\usepackage{amsmath}
\begin{document}

\begin{titlepage}
\begin{flushright}
CP3-21-09\\
\end{flushright}

\vspace{5pt}

\begin{center}

\vspace{7pt}

{\Large\bf Deformed Hopfion-Ra\~nada Knots in ModMax Electrodynamics}

\vspace{60pt}

Cl\'ementine Dassy$^{a}$ 
and Jan Govaerts$^{a,b,}$\footnote{
Fellow of the Stellenbosch Institute for Advanced Study (STIAS), Stellenbosch,
Republic of South Africa}

\vspace{30pt}

$^{a}${\sl Centre for Cosmology, Particle Physics and Phenomenology (CP3),\\
Institut de Recherche en Math\'ematique et Physique (IRMP),\\
Universit\'e catholique de Louvain (UCLouvain),\\
2, Chemin du Cyclotron, B-1348 Louvain-la-Neuve, Belgium}\\
{\ }\\
E-mail: {\em Clementine.Dassy@uclouvain.be, Jan.Govaerts@uclouvain.be}\\
\vspace{25pt}
$^{b}${\sl International Chair in Mathematical Physics and Applications (ICMPA--UNESCO Chair)\\
University of Abomey-Calavi, 072 B.P. 50, Cotonou, Republic of Benin}\\

\vspace{10pt}


\vspace{10pt}

\begin{abstract}
\noindent
Source-free so-called ModMax theories of nonlinear electrodynamics in the four dimensional Minkowski spacetime vacuum
are the only possible continuous deformations --- and as a function of a single real and positive parameter --- of source-free Maxwell linear electrodynamics
in the same vacuum, which preserve all the same Poincar\'e and conformal spacetime symmetries as well as the continuous
duality invariance of Maxwell's theory. Null field configurations of the latter however, including null electromagnetic knots, are singular for the Lagrangian formulation
of any spacetime Poincar\'e and conformal invariant theory of nonlinear electrodynamics. In particular null hopfion-Ra\~nada knots
are a distinguished and fascinating class on their own of topologically nontrivial solutions to Maxwell's equations. This work addresses the fate of these
configurations within ModMax theories. A doubled class of ModMax deformed hopfion-Ra\~nada knots is thereby identified, each of which
coalescing back in a continuous fashion to the original hopfion-Ra\~nada knot when the nonlinear deformation parameter is turned off.

\end{abstract}

\end{center}

\end{titlepage}

\setcounter{footnote}{0}

\section{Introduction}
\label{Intro}

Consider the equations of motion of source-free Maxwell Linear Electrodynamics (MLE) in the four dimensional Minkowski spacetime vacuum,
whether in spacetime covariant form,
\begin{equation}
\partial_\mu F^{\mu\nu}=0,\qquad
\partial_\mu{}^*F^{\mu\nu}=0,\qquad
F_{\mu\nu}=\partial_\mu A_\nu - \partial_\nu A_\mu,\qquad
^*F^{\mu\nu}=\frac{1}{2}\epsilon^{\mu\nu\rho\sigma} F_{\rho\sigma},
\end{equation}
or in 3-vector covariant form,
\begin{equation}
\vec{\nabla}\cdot\vec{E}=0,\qquad
\vec{\nabla}\times\vec{B}-\partial_t\vec{E}=\vec{0},\qquad
\vec{\nabla}\cdot\vec{B}=0,\qquad
\vec{\nabla}\times\vec{E}+\partial_t\vec{B}=\vec{0},
\end{equation}
in notations that are standard, $A^\mu(x^\mu)$ being the electromagnetic gauge field degrees of freedom\footnote{
Natural units such that $c=1$ and $\epsilon_0=1=\mu_0$ are in use with cartesian spacetime coordinates
$x^\mu=(t,x,y,z)$ ($\mu,\nu=0,1,2,3$), a mostly negative signature
for the Minkowski spacetime metric, ${\rm diag}\,\eta_{\mu\nu}=(+---)$, and $\epsilon^{0123}=+1$. In particular $F_{0i}=E^i=-\partial_i A^0-\partial_t A^i$,
$F_{ij}= -\epsilon^{ijk} B^k$ with $B^i=\epsilon^{ijk}\partial_j A^k$, and $\epsilon^{123}=+1$ with $i,j,k=1,2,3$, so that
$^*F_{0i}=B^i$ and $^*F_{ij}=\epsilon^{ijk}E^k$ (the summation convention is in place throughout).}.

It may possibly come as a surprise to learn that, even though these equations are intrinsically linear, they possess topologically nontrivial configurations
of electric and magnetic fields, known as electromagnetic knots\cite{Review1,Review2}, such that their closed electric and magnetic field lines are knotted
and linked with one another and maintain such topological structure throughout their time evolution. These solutions include null knots as a distinguished
subclass, namely such that $|\vec{E}|=|\vec{B}|$
and $\vec{E}\cdot\vec{B}=0$ everywhere in spacetime, that lend themselves to efficient methods of construction, based on the Bateman approach
with its pair of self-dual complex scalar potentials and involving the complex Riemann-Silberstein vector $\vec{R}=\vec{E} + i \vec{B}$.

Among these null knots the so-called hopfion-Ra\~nada (HR) knots\cite{HR1,HR2} stand out as remarkable configurations in a class of their own. Indeed
a HR knot may uniquely be constructed directly in terms of a self-dual doublet of Hopf fibrations
of the 3-sphere, one such fibration for each of the electric and magnetic components of the electromagnetic field. However a HR knot owes its existence not
only to this singular situation, but also to all the symmetries that the MLE equations possess, namely not only spacetime Poincar\'e covariance
but more largely spacetime conformal covariance specifically in 4 spacetime dimensions, and then global U(1) or SO(2) duality invariance as well that mixes
the electric and magnetic fields --- as well as $F_{\mu\nu}$ and $^*F_{\mu\nu}$ --- into one another in a continuous fashion (which is tantamount
to a simple global complex phase transformation of $\vec{R}$) . Indeed based on combinations of such symmetry transformations
and then including even the possibility of complex transformation parameters\cite{Review1,Review2}, it is possible to reach the HR knot starting
from a field configuration which may be even as trivial as that, say, of static and homogeneous null electric and magnetic fields.

As may be expected from topologically nontrivial field configurations the existence of these HR knots appears to be robust against any
nonlinear deformation of the source-free Maxwell equations in vacuum. The case has been made\cite{Goulart} that when considering the equations
of NonLinear Electrodynamics (NLE) which are deformations of the equations of MLE, any exact null knot solution of MLE remains an exact solution
of NLE, at least provided Poincar\'e covariance is preserved in the deformed NLE. The argument is straightforward. Any gauge and Poincar\'e invariant
quantity built out of the electromagnetic vector potential $A^\mu$ and its field strength $F_{\mu\nu}$ alone may be constructed out of only\cite{Escobar}
the following two well-known gauge invariant and Lorentz scalar or pseudo-scalar quantities,
\begin{equation}
{\cal S}=-\frac{1}{4}F_{\mu\nu}{}F^{\mu\nu}=\frac{1}{2}\left(\vec{E}^2 - \vec{B}^2\right),\qquad
{\cal P}=-\frac{1}{4}F_{\mu\nu}{}^*F^{\mu\nu}=\vec{E}\cdot\vec{B}.
\end{equation}
In other words, within the Lagrangian formulation, through the variational principle any Poincar\'e invariant nonlinear electrodynamics with second order
in time only equations of motion derives from a Lagrangian density which is some given but otherwise arbitrary function of these two gauge and Lorentz
invariant quantities, ${\cal L}({\cal S},{\cal P})$ (provided ${\cal L}({\cal S},{\cal P})$
is even in ${\cal P}$, parity invariance is preserved as well). The case of MLE simply corresponds to ${\cal L}_0({\cal S},{\cal P})={\cal S}$.
Consider then the associated equations of motion, whether in spacetime covariant form,
\begin{equation}
\partial_\mu G^{\mu\nu}=0,\qquad
\partial_\mu{}^*F^{\mu\nu}=0,\qquad
G_{\mu\nu}={\cal L}_{\cal S} F_{\mu\nu}\,+\,{\cal L}_{\cal P}{}^*F_{\mu\nu},\qquad
^*G^{\mu\nu}=\frac{1}{2}\epsilon^{\mu\nu\rho\sigma}G_{\rho\sigma},
\label{eq:COV}
\end{equation}
or in 3-vector covariant form,
\begin{equation}
\vec{\nabla}\cdot\vec{D}=0,\qquad
\vec{\nabla}\times\vec{H}-\partial_t\vec{D}=\vec{0},\qquad
\vec{\nabla}\cdot\vec{B}=0,\qquad
\vec{\nabla}\times\vec{E}+\partial_t\vec{B}=\vec{0},
\end{equation}
where ${\cal L}_{\cal S}=\partial_{\cal S}{\cal L}({\cal S},{\cal P})$, ${\cal L}_{\cal P}=\partial_{\cal P}{\cal L}({\cal S},{\cal P})$, while\footnote{Note
that  for MLE with ${\cal L}({\cal S},{\cal P})={\cal L}_0({\cal S})={\cal S}$, $\vec{D}=\vec{E}$, $\vec{H}=\vec{B}$, and $G_{\mu\nu}=F_{\mu\nu}$.},
\begin{equation}
\vec{D}=\frac{\partial{\cal L}}{\partial\vec{E}}={\cal L}_{\cal S}\,\vec{E}\,+\,{\cal L}_{\cal P}\,\vec{B},\qquad
\vec{H}=-\frac{\partial{\cal L}}{\partial\vec{B}}={\cal L}_{\cal S}\,\vec{B}\,-\,{\cal L}_{\cal P}\,\vec{E},
\label{eq:DH}
\end{equation}
which are such that $G_{0i}=D^i$ and $G_{ij}=-\epsilon^{ijk}H^k$, as well as $^*G_{0i}=H^i$ and $^*G_{ij}=\epsilon^{ijk}D^k$.
Clearly then any exact null field solution to the MLE equations of motion remains an exact null field solution to the NLE equations of motion,
since then one has ${\cal S}=0={\cal P}$ so that the coefficients ${\cal L}_{\cal S}$ and ${\cal L}_{\cal P}$ are then constant coefficients
in the above NLE equations. Consequently any null knot solution for MLE, in particular the HR one, remains a solution for NLE.

This argument however, presents a loop-hole in that it applies provided only that the quantities ${\cal L}_{\cal S}$ and ${\cal L}_{\cal P}$
be well-defined at $({\cal S},{\cal P})=(0,0)$, namely provided that the function ${\cal L}({\cal S},{\cal P})$ be analytic in both its variables
at $({\cal S},{\cal P})=(0,0)$. Obviously this leaves out all conformally invariant NLE theories. Indeed the Lagrangian
density of any conformally invariant NLE deformation of MLE is necessarily of the form
\begin{equation}
{\cal L}({\cal S},{\cal P})={\cal S}\cdot F\left(\frac{\cal P}{{\cal S}}\right),
\end{equation}
where $F(u)$ is some given but otherwise arbitrary function of a single variable $u$ (to be chosen so that restrictions of classical causality and quantum
unitarity be met as well\cite{Shabad,Denisov}). Clearly unless there is no dependency on ${\cal P}$ at all --- which
would lead back to MLE --- any such Lagrangian density is nonanalytic at $({\cal S},{\cal P})=(0,0)$. In such instance the fate of the null solutions
of MLE, in particular of null knots inclusive of the HR one, remains an open question for whatever conformally invariant NLE theory.

Given the role played by all symmetries of MLE for the existence of null HR knots, one may wish to retain as much as may be feasible all those same
symmetries for a NLE theory. Besides the usual MLE theory, a recent result\cite{Town} has established that there exist only two other possible NLE theories
which preserve exactly all the same symmetries of spacetime Poincar\'e and conformal invariance and of electromagnetic duality between
the electric and magnetic sectors. One of these NLE theories, dubbed BB electrodynamics by the authors of Ref.\cite{Town}
(for Bialynicki-Birula\cite{BB}), corresponds to a Hamiltonian density given as ${\cal H}_{BB}=|\vec{D}\times\vec{B}|$.
The other, dubbed ModMax theories by their discoverers\cite{Town},
corresponds to a continuous deformation of MLE parametrised by a single real and positive parameter, $\gamma\ge 0$,
which reduces to MLE for $\gamma=0$, and of which the Lagrangian density is uniquely given by\cite{Town,Kosy},
\begin{eqnarray}
{\cal L}_\gamma({\cal S},{\cal P}) &=& \cosh\gamma\,{\cal S}\,+\,\sinh\gamma\,\sqrt{{\cal S}^2+{\cal P}^2} \nonumber \\
&=& \frac{1}{2}\cosh\gamma\left(\vec{E}^2-\vec{B}^2\right)+\frac{1}{2}\sinh\gamma\,
\sqrt{\left(\vec{E}^2-\vec{B}^2\right)^2+4\left(\vec{E}\cdot\vec{B}\right)^2}.
\end{eqnarray}
As is characteristic of NLE theories\cite{BB,Gibbons,Deser1,Deser2}, the Poincar\'e and conformal invariance of ModMax dynamics is manifest within
the Lagrangian formulation but not its duality invariance, while the latter is manifest within the Hamiltonian formulation but then not its Poincar\'e
and conformal invariance.

One main purpose of the present work is to better understand which fate awaits the null HR knot when the ModMax parameter $\gamma$
is turned on. Would it still remain a solution as such or would it need to be deformed continuously into some other configuration in order to remain
a solution to the ModMax equations of motion, and if yes, how would all this work out? As we shall establish, all these questions do have
an answer, but may be then in a somewhat surprising way, and then leading to still further questions yet to be unravelled.

Our discussion is organised as follows. Sect.\ref{Sect2} first presents a review of the Lagrangian and Hamiltonian formulations for any nonlinear
electrodynamics theory, and then addresses the properties implied by spacetime Poincar\'e invariance or duality invariance which are best
represented by introducing a generalisation of the Riemann-Silberstein vector to the nonlinear context and an associated complex 4-vector potential.
Finally that same Section considers some methods for the construction of solutions and in particular, by providing a generalisation as well of the usual Bateman
approach with its complex scalar potentials. In Sect.\ref{Sect3} the discussion is restricted to ModMax theories specifically, first by outlining some general remarks
related to the construction of classes of solutions, and then by presenting two explicit classes of solutions based on the generalised Bateman approach.
Then finally it is Sect.\ref{Sect4} that achieves the explicit analytic construction of two new classes of electromagnetic knots solving the ModMax equations,
which each are continuous deformations of the ordinary hopfion-Ra\~nada knot, thereby retaining the topologically nontrivial structures of the latter.
Concluding comments are presented in a last Section.

\section{Nonlinear Electrodynamics Theories}
\label{Sect2}

\subsection{Lagrangian and Hamiltonian formulations}
\label{Sect2.1}

Before embarking on the study of ModMax theories {\sl per se}, let us first consider an arbitrary NLE theory\cite{BB} with second order in time only equations
of motion, and both its Lagrangian and Hamiltonian (or first-order) formulations. At this stage the theory need not be spacetime covariant or even 3d covariant,
but only gauge invariant under gauge transformations of the electromagnetic 4-vector potential $A^\mu(x^\mu)$, which constitutes the actual degrees
of freedom of the Lagrangian formulation.

Within the latter formulation and because of gauge invariance the corresponding Lagrangian density is then necessarily some given but otherwise arbitrary function
of $F_{\mu\nu}=\partial_\mu A_\nu - \partial_\nu A_\mu$ or equivalently of $\vec{E}=-\vec{\nabla}A^0-\partial_t \vec{A}$ and $\vec{B}=\vec{\nabla}\times\vec{A}$,
namely,
\begin{equation}
{\cal L}_{NLE}(F_{\mu\nu})={\cal L}_{NLE}(\vec{E},\vec{B})={\cal L}_{NLE}(-\vec{\nabla}A^0-\partial_t\vec{A},\vec{\nabla}\times\vec{A}),
\end{equation}
thus indeed a function of $(A^\nu,\partial_\mu A^\nu)$, $A^\nu$ being the actual fundamental configuration space degrees of freedom for this formulation.
Note well that at this stage the Lagrangian density need not yet be Poincar\'e invariant, and not even 3d rotational invariant, in spite of the
3-vector notation being used. However if 3d rotational invariance is implemented the present notation makes this 3d covariance under spatial rotations 
manifest (as would the spacetime covariant notation in case of complete Poincar\'e invariance).

Because of the Bianchi identity that applies to $F_{\mu\nu}$ constructed in terms of $A^\mu$, namely $\partial_\mu{}^*F^{\mu\nu}=0$,
the Lagrangian equations of motion for the electric and magnetic fields include the usual two homogeneous Maxwell equations
(for $\nu=0$ and $\nu=i$, respectively),
\begin{equation}
\vec{\nabla}\cdot\vec{B}=0,\qquad
\vec{\nabla}\times\vec{E}+\partial_t\vec{B}=\vec{0}.
\label{eq:NLE1}
\end{equation}
The Euler-Lagrange equations of motion for $A^0$ and $\vec{A}$, respectively, deriving from the considered Lagrangian are,
\begin{equation}
\vec{\nabla}\cdot\vec{D}=0,\qquad
\vec{\nabla}\times\vec{H}-\partial_t\vec{D}=\vec{0},
\label{eq:NLE2}
\end{equation}
with the definitions
\begin{equation}
\vec{D}(\vec{E},\vec{B})=\frac{\partial{\cal L}_{NLE}}{\partial\vec{E}}(\vec{E},\vec{B}),\qquad
\vec{H}(\vec{E},\vec{B})=-\frac{\partial{\cal L}_{NLE}}{\partial\vec{B}}(\vec{E},\vec{B}),
\label{eq:Const}
\end{equation}
it being understood that the gauge invariant fields $(\vec{E},\vec{B})$ derive from the gauge variant ones $(A^0,\vec{A}\,)$ through the relations recalled above.
The NLE equations of motion are thus given in (\ref{eq:NLE1}) and (\ref{eq:NLE2}) with the constitutive equations (\ref{eq:Const}).

In order to identify the corresponding first-order (or Hamiltonian) formulation, in addition to the gauge field $A^\mu(x^\mu)$ let us introduce furthermore
an independent antisymmetric field $F_{\mu\nu}(x^\mu)=-F_{\nu\mu}(x^\mu)$ (not yet related in any way to the curl of $A_\mu$),
and consider the following first-order Lagrangian density constructed out of ${\cal L}_{NLE}(F_{\mu\nu})$ above\footnote{The factor $1/2$
in the first term on the r.h.s.~of this expression accounts for the fact that when applying to it the variational principle the degrees of freedom
$F_{\mu\nu}$ and $F_{\nu\mu}$ for $\mu\ne \nu$ are not to be considered as being independent but rather to be related by $F_{\nu\mu}=-F_{\mu\nu}$.}\cite{Gibbons},
\begin{equation}
{\cal L}^{(1)}_{NLE}(A^\mu;F_{\mu\nu})=-\frac{1}{2}\frac{\partial{\cal L}_{NLE}(F_{\mu\nu})}{\partial F_{\mu\nu}}
\left[F_{\mu\nu}-(\partial_\mu A_\nu - \partial_\nu A_\mu)\right]\,+\,{\cal L}_{NLE}(F_{\mu\nu}).
\end{equation}
Provided the Hessian of the function ${\cal L}_{NLE}(F_{\mu\nu})$ relative to the variables $F_{\mu\nu}$ is regular --- a property which is
assumed implicitly throughout ---, the Euler-Lagrange equations for $F_{\mu\nu}$ readily reduce to
\begin{equation}
F_{\mu\nu}=\partial_\mu A_\nu - \partial_\nu A_\mu.
\end{equation}
It is thereby established that the dynamics deriving from both Lagrangian densities are equivalent, however with ${\cal L}^{(1)}_{NLE}$
being first-order in time derivatives (of $A^\mu$) and thus Hamiltonian.

In order to display this Hamiltonian structure let us make explicit the contributions to ${\cal L}^{(1)}_{NLE}$ in the fields
$E^i=F_{0i}$ and $B^i=-\epsilon^{ijk}F_{jk}$, and use the definitions in (\ref{eq:Const}), so that,
\begin{eqnarray}
\label{eq:1storder}
&&{\cal L}^{(1)}_{NLE}(A^0,\vec{A};\vec{E},\vec{B})= \nonumber \\
&& = -\vec{D}\cdot\left(\vec{E}+\partial_t \vec{A}+\vec{\nabla}A^0\right) + \vec{H}\cdot\left(\vec{B}-\vec{\nabla}\times\vec{A}\right) + {\cal L}_{NLE}(\vec{E},\vec{B})  \\
&& = -\partial_t\vec{A}\cdot\vec{D}-\left(\vec{E}\cdot\vec{D}-{\cal L}_{NLE}(\vec{E},\vec{B})\right) + A^0\vec{\nabla}\cdot\vec{D}
 - \vec{\nabla}\cdot(A^0\vec{D}) + \vec{H}\cdot\left(\vec{B}-\vec{\nabla}\times\vec{A}\right) . \nonumber
\end{eqnarray}
Clearly and as is well known\cite{Gov} the $A^0$ component of the gauge field $A^\mu$ is seen to play the role of a Lagrange multiplier for the first-class constraint
$\vec{\nabla}\cdot\vec{D}=0$, namely Gauss' law, while the term $-\vec{\nabla}\cdot(A^0\vec{D})$ is a spacelike surface term which at infinity
does not contribute to the total action (when assuming sufficient rapid fall-off of the field $\vec{D}$). However both terms
$A^0\vec{\nabla}\cdot\vec{D}-\vec{\nabla}\cdot(A^0\vec{D})=-\vec{\nabla}A^0\cdot\vec{D}$ will be kept in the latter form, for later purposes.
On the other hand the variation of the corresponding action with respect to $\vec{B}$ (and accounting for the dependency
of $\vec{H}$ on $\vec{B}$ such that the Hessian of ${\cal L}_{NLE}(\vec{E},\vec{B})$ relative to $\vec{B}$ be regular) implies once again the equation
$\vec{B}=\vec{\nabla}\times\vec{A}$, of which the solution may be used as such by substitution into the above expression, so that the term
in $\vec{H}$ no longer contributes to the first-order Lagrangian density\footnote{By considering
from the outset an expression for ${\cal L}^{(1)}_{NLE}$ which does not include the second term in $\vec{H}$ on the r.h.s.~of (\ref{eq:1storder})
one readily reaches the same conclusion for the first-order formulation.}. Furthermore the second term in parentheses is an invitation to consider the
following Legendre transformation of ${\cal L}_{NLE}(\vec{E},\vec{B})$ relative to $\vec{E}$, which defines the first-class NLE Hamiltonian density
${\cal H}_{NLE}$,
\begin{equation}
{\cal H}_{NLE}(\vec{D},\vec{B})=\vec{E}\cdot\vec{D}-{\cal L}_{NLE}(\vec{E},\vec{B}),
\label{eq:Legendre1}
\end{equation}
such that,
\begin{equation}
\vec{D}(\vec{E},\vec{B})=\frac{\partial{\cal L}_{NLE}(\vec{E},\vec{B})}{\partial\vec{E}},\qquad
\vec{E}(\vec{D},\vec{B})=\frac{\partial{\cal H}_{NLE}(\vec{D},\vec{B})}{\partial\vec{D}},
\label{eq:Legendre2}
\end{equation}
while
\begin{equation}
\vec{H}=-\frac{\partial{\cal L}_{NLE}(\vec{E},\vec{B})}{\partial\vec{B}}=\frac{\partial{\cal H}_{NLE}(\vec{D},\vec{B})}{\partial\vec{B}}.
\end{equation}
And finally, wanting to make explicit a symmetry between the electric and magnetic sectors of the action, let us introduce an additional Lagrange
multiplier, $C^0$, playing a role analogous to that of $A^0$, but this time for the constraint $\vec{\nabla}\cdot\vec{B}=0$
with $\vec{B}=\vec{\nabla}\times\vec{A}$, by adding to this Lagrangian density an extra term given by $\vec{\nabla}C^0\cdot(\vec{\nabla}\times\vec{A})$
(which in effect is a spacelike surface term on its own, but again useful for later purposes).

Hence in conclusion the first-order Hamiltonian action density is given in the form
\begin{equation}
{\cal L}^{(1)}_{NLE}(\vec{A},\vec{D};A^0, C^0)=-\partial_t\vec{A}\cdot\vec{D}-{\cal H}_{\rm total}(\vec{D},\vec{\nabla}\times\vec{A};A^0, C^0),
\end{equation}
with the total first-class Hamiltonian density
\begin{equation}
{\cal H}_{\rm total}(\vec{D},\vec{\nabla}\times\vec{A};A^0,C^0)={\cal H}_{NLE}(\vec{D},\vec{\nabla}\times\vec{A}) + \vec{\nabla}A^0\cdot\vec{D}
-\vec{\nabla}C^0\cdot\left(\vec{\nabla}\times\vec{A}\right),
\end{equation}
and a phase space spanned by the fields $(\vec{A}(t,\vec{x}),\vec{D}(t,\vec{x}))$.
In particular the very first contribution on the r.h.s~of this final expression for ${\cal L}^{(1)}_{NLE}(\vec{A},\vec{D};A^0)$ shows that the pairs of components
$(A^i,-D^i)$ are canonically conjugate variables for each separate value $i=1,2,3$. Computing the relevant Poisson brackets with the total Hamiltonian
it then follows that the Hamiltonian equation of motion for $\vec{D}$ reads
\begin{equation}
\partial_t\vec{D}=\vec{\nabla}\times\vec{H},\qquad
\vec{\nabla}\times\vec{H}-\partial_t\vec{D}=\vec{0},
\end{equation}
while that for $\vec{A}$ leads to
\begin{equation}
\partial_t\vec{A}=-\vec{E}-\vec{\nabla}A^0,\qquad
\vec{E}=-\partial_t\vec{A}-\vec{\nabla}A^0,
\end{equation}
which in turn implies, using $\vec{\nabla}\times\vec{A}=\vec{B}$,
\begin{equation}
\vec{\nabla}\times\vec{E}+\partial_t\vec{B}=\vec{0}.
\end{equation}
This set of Hamiltonian equations of motion remains to be supplemented with the first-class constraint of Gauss' law, which is implied
by the variation relative to the Lagrange multiplier $A^0$, namely
\begin{equation}
\vec{\nabla}\cdot\vec{D}=0,
\end{equation}
while finally one has as well from the variation with respect to the Lagrange multiplier $C^0$, in a manner consistent with the fact that
$\vec{B}=\vec{\nabla}\times\vec{A}$,
\begin{equation}
\vec{\nabla}\cdot\vec{B}=0.
\end{equation}
Exactly the same set of NLE equations of motion as in (\ref{eq:NLE1}) and (\ref{eq:NLE2}) for the fields $(\vec{E},\vec{D},\vec{B},\vec{H})$
is thus recovered, with this time the fields
$(\vec{D},\vec{B}=\vec{\nabla}\times\vec{A})$ as the (fundamental) phase space degrees of freedom, while the derived fields $(\vec{E},\vec{H})$
are determined through the constitutive equations,
\begin{equation}
\vec{E}(\vec{D},\vec{B})=\frac{\partial{\cal H}_{NLE}(\vec{D},\vec{B})}{\partial\vec{D}},\qquad
\vec{H}(\vec{D},\vec{B})=\frac{\partial{\cal H}_{NLE}(\vec{D},\vec{B})}{\partial\vec{B}},
\label{eq:Const2}
\end{equation}
where ${\cal H}_{NLE}(\vec{D},\vec{B})$ is the Legendre transform of ${\cal L}_{NLE}(\vec{E},\vec{B})$ relative to $\vec{E}$.

It turns out that the Hamiltonian formulation of such nonlinear electrodynamics is best suited in order to solve these equations
in view of the purposes of the present work, namely the construction of electromagnetic knots.

Further considerations remain in order however, to make the electric-magnetic symmetry as explicit as feasible.
Note that in the same way that the equation $\vec{\nabla}\cdot\vec{B}=0$ is solved
in terms of the magnetic vector potential $\vec{A}$ and $\vec{B}=\vec{\nabla}\times\vec{A}$, Gauss' law $\vec{\nabla}\cdot\vec{D}=0$ may be solved
in terms of an electric vector potential $\vec{C}$ such that $\vec{D}=\vec{\nabla}\times\vec{C}$. The identification of both these vector potentials is
then defined up to the spatial gradient of a different arbitrary scalar field in each case, namely a double gauge invariance to be addressed hereafter.
The two 3-vector equations remaining to be considered are then,
\begin{equation}
\vec{\nabla}\times\left(\vec{E}+\partial_t\vec{A}\right)=\vec{0},\qquad
\vec{\nabla}\times\left(\vec{H}-\partial_t\vec{C}\right)=\vec{0},
\end{equation}
while the first-order action density is then expressed as\footnote{Each of the terms involving $A^0$ and $C^0$ are thus in themselves spacelike surface terms,
retained here in the local first-order Lagrangian density, hence contributing to the local equations of motion.},
\begin{eqnarray}
{\cal L}^{(1)}_{NLE}(\vec{A},\vec{C};A^0,C^0) \!\!\! &=& \!\!\! -\partial_t\vec{A}\cdot\left(\vec{\nabla}\times\vec{C}\right) - \nonumber \\
&&\!\!\! -\left[{\cal H}_{NLE}(\vec{\nabla}\times\vec{C},\vec{\nabla}\times\vec{A})
+\vec{\nabla}A^0\cdot\left(\vec{\nabla}\times\vec{C}\right)-\vec{\nabla}C^0\cdot\left(\vec{\nabla}\times\vec{A}\right)\right].
\label{eq:L1}
\end{eqnarray}
In this formulation phase space is spanned by the fields $(\vec{A},\vec{C})$, however now with a noncanonical symplectic structure implicitly defined
by the very first term in the r.h.s~of this expression. Their equations of motion are then given in the form,
\begin{equation}
\partial_t\vec{A}=-\vec{E}-\vec{\nabla}A^0,\qquad
\partial_t\vec{C}=\vec{H}-\vec{\nabla}C^0.
\end{equation}
Even though the two remaining 3-vector equations for $\vec{E}$ and $\vec{H}$
may be solved in terms of the two scalar fields $A^0$ and $C^0$, such that
\begin{equation}
\vec{E}=-\partial_t\vec{A}-\vec{\nabla}A^0,\qquad
\vec{H}=\partial_t\vec{C}+\vec{\nabla}C^0,
\end{equation}
the two sets of fields $(A^0,\vec{A})$ and $(C^0,\vec{C})$ are constrained to satisfy the two constitutive equations in (\ref{eq:Const2})
with $\vec{D}=\vec{\nabla}\times\vec{C}$ and $\vec{B}=\vec{\nabla}\times\vec{A}$.

Using for half of the first term on the r.h.s.~of (\ref{eq:L1}) the following identity valid for any two vector fields $\vec{V}(x^\mu)$ and $\vec{W}(x^\mu)$,
\begin{equation}
\partial_t\vec{V}\cdot\left(\vec{\nabla}\times\vec{W}\right)=
\partial_t\left[\vec{V}\cdot\left(\vec{\nabla}\times\vec{W}\right)\right]
+\vec{\nabla}\cdot\left(\vec{V}\times\partial_t\vec{W}\right) - \partial_t\vec{W}\cdot\left(\vec{\nabla}\times\vec{V}\right),
\end{equation}
while discarding the related time- and space-like surface terms at infinity,
and given the Hamiltonian density ${\cal H}_{NLE}(\vec{D},\vec{B})$, finally the first-order Hamiltonian action of such NLE theory is expressed as,
\begin{eqnarray}
\hspace{-15pt}S^{(1)}_{NLE}[\vec{A},\vec{C};A^0,C^0\,]\!\!\! &=&\!\!\! \int_{(\infty)}d^4x^\mu\left\{
-\frac{1}{2}\partial_t\vec{A}\cdot\left(\vec{\nabla}\times\vec{C}\right) + \frac{1}{2}\partial_t\vec{C}\cdot\left(\vec{\nabla}\times\vec{A}\right)\,-\,\right. \nonumber \\
&&\left. \,-\,\left[{\cal H}_{NLE}(\vec{\nabla}\times\vec{C},\vec{\nabla}\times\vec{A})
+\vec{\nabla}A^0\cdot\left(\vec{\nabla}\times\vec{C}\right)-\vec{\nabla}C^0\cdot\left(\vec{\nabla}\times\vec{A}\right)\right] \right\}.
\label{eq:S1-NLE}
\end{eqnarray}
This Hamiltonian formulation of nonlinear electrodynamics is thus in direct correspondence with its Lagrangian one in the form of
\begin{equation}
S_{NLE}[A^0,\vec{A}\,]=\int_{(\infty)}d^4x^\mu\,{\cal L}_{NLE}(-\vec{\nabla}A^0-\partial_t\vec{A},\vec{\nabla}\times\vec{A}),
\end{equation}
provided the Legendre transformation in (\ref{eq:Legendre1}) and (\ref{eq:Legendre2}) between ${\cal L}_{NLE}(\vec{E},\vec{B})$
and ${\cal H}_{NLE}(\vec{D},\vec{B})$ be well defined.

To conclude let us address the double local gauge invariance of this first-order Hamiltonian formulation of NLE theories. Given any two arbitrary
real scalar fields $\varphi_e(x^\mu)$ and $\varphi_m(x^\mu)$, it may readily be checked that the action $S^{(1)}_{NLE}[\vec{A},\vec{C};A^0,C^0\,]$
is invariant up to a spacelike surface term under the following transformations of the pairs of scalar and vector potentials
$(A^0,\vec{A})$ and $(C^0,\vec{C})$,
\begin{equation}
{C^0}'=C^0+\partial_t\varphi_e,\qquad
\vec{C}'=\vec{C}-\vec{\nabla}\varphi_e ;\qquad
{A^0}'=A^0+\partial_t\varphi_m,\qquad
\vec{A}'=\vec{A}-\vec{\nabla}\varphi_m,
\end{equation}
which thus define the double local gauge invariance of the Hamiltonian formulation of any source-free NLE theory in the Minkowski spacetime vacuum.

\subsection{Poincar\'e and duality invariances, and the Riemann-Silberstein tensor}
\label{Sect2.2}

From here on let us assume that the NLE theory under consideration is Poincar\'e invariant, namely with
\begin{equation}
{\cal L}_{NLE}(F_{\mu\nu})={\cal L}_{NLE}(\vec{E},\vec{B})={\cal L}({\cal S},{\cal P})={\cal L}\left(\frac{1}{2}(\vec{E}^2-\vec{B}^2),\vec{E}\cdot\vec{B}\right).
\end{equation}
This implies that it ought to be possible to give its equations of motion a manifest spacetime covariant form, at least within the Lagrangian formulation.
Covariance under 3d rotations is readily manifest, in both formulations, given the 3-vector form already given above to the relevant equations of motion.
However covariance under Lorentz boosts cannot be made manifest within the Hamiltonian formulation. Nevertheless the action of these symmetry
transformations on the different fields involved may be identified, based on the Lagrangian formulation.

Consider a Lorentz boost of 3-velocity vector $\vec{\beta}_0\ne \vec{0}$, with the associated Lorentz dilation factor $\gamma_0=(1-\vec{\beta}^2_0)^{-1/2}$
and unitary boost direction $\hat{\beta}_0=\vec{\beta}_0/\beta_0$ with $\beta_0=|\vec{\beta}_0|$.
The 4-vector of spacetime coordinates $x^\mu=(t,\vec{x})$ then transforms according to the relations,
\begin{equation}
x'^{\mu}=(t',\vec{x}\,'):\qquad t'=\gamma_0(t-\vec{\beta}_0\cdot\vec{x}),\qquad
\vec{x}\,'=\vec{x}+\hat{\beta}_0\left[-\beta_0\gamma_0 t + (\gamma_0-1)\hat{\beta}_0\cdot\vec{x}\right].
\end{equation}
The Lorentz boost transformation properties of $F_{\mu\nu}$ and its components $(\vec{E},\vec{B})$ readily follow from the spacetime covariant
properties of the 4-vector $A^\mu$. As is well known one finds,
\begin{equation}
\vec{E}\,'=\gamma_0(\vec{E}+\vec{\beta}_0\times\vec{B})-(\gamma_0-1)(\hat{\beta}_0\cdot\vec{E})\,\hat{\beta}_0,
\end{equation}
\begin{equation}
\vec{B}'=\gamma_0(\vec{B}-\vec{\beta}_0\times\vec{E})-(\gamma_0-1)(\hat{\beta}_0\cdot\vec{B})\,\hat{\beta}_0.
\end{equation}
As a consequence, and given that ${\cal L}_{\cal S}({\cal S},{\cal P})$ and ${\cal L}_{\cal P}({\cal S},{\cal P})$ are Poincar\'e invariant for a Poincar\'e invariant NLE,
based on the definitions (\ref{eq:DH}) it follows that the fields $(\vec{D},\vec{H})$ possess the same transformation properties
as $(\vec{E},\vec{B})$ do, namely,
\begin{equation}
\vec{D}\,'=\gamma_0(\vec{D}+\vec{\beta}_0\times\vec{H})-(\gamma_0-1)(\hat{\beta}_0\cdot\vec{D})\,\hat{\beta}_0,
\end{equation}
\begin{equation}
\vec{H}'=\gamma_0(\vec{H}-\vec{\beta}_0\times\vec{D})-(\gamma_0-1)(\hat{\beta}_0\cdot\vec{H})\,\hat{\beta}_0.
\end{equation}
Of course this is obviously consistent with the NLE equations of motion, which are thus covariant under Lorentz boosts as well as under 3d rotations,
\begin{equation}
\vec{\nabla}\cdot\vec{B}=0,\quad
\vec{\nabla}\times\vec{E}+\partial_t\vec{B}=\vec{0};\qquad
\vec{\nabla}\cdot\vec{D}=0,\quad
\vec{\nabla}\times\vec{H}-\partial_t\vec{D}=\vec{0}.
\end{equation}
Note that the manifest spacetime covariance of these equations is made explicit already in the form of (\ref{eq:COV}) within the Lagrangian formalism,
in terms of the two anti-symmetric covariant tensors $F_{\mu\nu}$ and $G_{\mu\nu}$ of which the independent components are the pairs of 3-vectors
$(\vec{E},\vec{B})$ and $(\vec{D},\vec{H})$, respectively. However this Poincar\'e covariance is not manifest from the Hamiltonian density ${\cal H}(\vec{D},\vec{B})$
obtained through the Legendre transformation in $\vec{E}$ of the Lagrangian density ${\cal L}({\cal S},{\cal P})$ under consideration,
\begin{equation}
{\cal H}(\vec{D},\vec{B})=\vec{E}\cdot\vec{D}-{\cal L}({\cal S},{\cal P}),\qquad
\vec{D}=\frac{\partial{\cal L}}{\partial\vec{E}},\qquad
\vec{E}=\frac{\partial{\cal H}}{\partial\vec{D}}.
\end{equation}

Given the central role played by the fields $(\vec{D},\vec{B})$ within the Hamiltonian formulation let us introduce a generalisation
of the complex Riemann-Silberstein vector usually considered for MLE\cite{Review1,Review2}\footnote{In the case of MLE since one has $\vec{D}=\vec{E}$
and $\vec{H}=\vec{B}$, hence $\vec{S}=-i\vec{R}$, there exists then a single independent Riemann-Silberstein vector.}, namely through a pair of independent
complex Riemann-Silberstein (RS) 3-vectors $\vec{R}$ and $\vec{S}$ defined as follows\footnote{As a matter of fact such definitions are relevant
independently of whether Poincar\'e or even only 3d rotational covariance is in place or not; these two vectors could already have been considered
in Sect.\ref{Sect2.1}.},
\begin{equation}
\vec{R}=\vec{D} + i \vec{B},\qquad
\vec{S}=\vec{H}-i\vec{E}=\frac{\partial{\cal H}(\vec{D},\vec{B})}{\partial\vec{B}} - i \frac{\partial{\cal H}(\vec{D},\vec{B})}{\partial\vec{D}}.
\label{eq:EOM-RS}
\end{equation}
In terms of these complex vectors the NLE equations of motion take the more compact form,
\begin{equation}
\vec{\nabla}\cdot\vec{R}=0,\qquad
\vec{\nabla}\times\vec{S}-\partial_t\vec{R}=\vec{0}.
\label{eq:EOM-RS2}
\end{equation}
Note how all nonlinearities reside in the sole occurrence in these expressions of the RS vector $\vec{S}$, through its dependence
on $(\vec{D},\vec{B})$ given its definition in (\ref{eq:EOM-RS}) involving ${\cal H}(\vec{D},\vec{B})$.

While the 3d rotational covariance of both $\vec{R}$ and $\vec{S}$ is manifest, from the above  transformations under Lorentz boosts
of the vectors $(\vec{E},\vec{B},\vec{D},\vec{H})$ one finds,
\begin{equation}
\vec{R}\,'=\gamma_0(\vec{R}+\vec{\beta}_0\times\vec{S})-(\gamma_0-1)(\hat{\beta}_0\cdot\vec{R})\,\hat{\beta}_0,
\end{equation}
\begin{equation}
\vec{S}'=\gamma_0(\vec{S}-\vec{\beta}_0\times\vec{R})-(\gamma_0-1)(\hat{\beta}_0\cdot\vec{S})\,\hat{\beta}_0.
\end{equation}
Thus under the full Lorentz group of spacetime symmetry transformations (3d rotations and Lorentz boosts),
the RS vectors $(\vec{R},\vec{S})$ and their components are mixed into one another in precisely the same way that
$(\vec{E},\vec{B})$ on the one hand, and $(\vec{D},\vec{H})$ on the other are. By analogy with $F_{\mu\nu}$ and its $(\vec{E},\vec{B})$ components,
this observation suggests to introduce the following 2-index antisymmetric spacetime covariant Riemann-Silberstein tensor $R_{\mu\nu}=-R_{\nu\mu}$ and
its dual $^*R_{\mu\nu}=\frac{1}{2}\epsilon_{\mu\nu\rho\sigma}R^{\rho\sigma}$ such that,
\begin{equation}
R_{0i}=R^i,\qquad R_{ij}=-\epsilon^{ijk}S^k,\qquad
^*R_{0i}=S^i,\qquad ^*R_{ij}=\epsilon^{ijk}R^k.
\end{equation}
In actual fact, in terms of the spacetime covariant tensors $F_{\mu\nu}$ and $G_{\mu\nu}$ introduced in (\ref{eq:COV}), one has,
\begin{equation}
R_{\mu\nu}=G_{\mu\nu} + i {}\,^*F_{\mu\nu},\qquad
^*R_{\mu\nu}=-i\left(F_{\mu\nu} + i{}\, ^*G_{\mu\nu}\right),
\end{equation}
thus confirming at once the spacetime covariance properties of $R_{\mu\nu}$. And in particular in manifestly spacetime covariant form
the full set of NLE equations of motion are now simply expressed as,
\begin{equation}
\partial_\mu R^{\mu\nu}=0,
\end{equation}
with the components $\nu=0$ and $\nu=i$ corresponding to the two equations in (\ref{eq:EOM-RS2}) in the same order, respectively.
However note that $\partial_\mu{}\,^*R^{\mu\nu}$ is not restricted in any particular way in NLE, since one finds,
\begin{equation}
\partial_\mu{}\,^*R^{\mu0}=\vec{\nabla}\cdot\vec{S},\qquad
\partial_\mu{}\,^*R^{\mu i}=-\left(\vec{\nabla}\times\vec{R}+\partial_t\vec{S}\right)^i.
\end{equation}

This last remark is to be contrasted with the situation for the specific case of MLE, corresponding to ${\cal L}_0={\cal S}=(\vec{E}^2-\vec{B}^2)/2$,
${\cal H}_0=(\vec{E}^2+\vec{B}^2)/2$, $\vec{D}=\vec{E}$, $\vec{H}=\vec{B}$, and thus $G_{\mu\nu}=F_{\mu\nu}$. Consequently one then has,
\begin{equation}
{\rm MLE}:\qquad R_{\mu\nu}=F_{\mu\nu} +i{}\,^*F_{\mu\nu},\qquad ^*R_{\mu\nu}=-i\, R_{\mu\nu},
\end{equation}
which implies that the source-free Maxwell equations are expressed by both the following equations,
\begin{equation}
{\rm MLE}:\qquad \partial_\mu R^{\mu\nu}=0,\qquad
\partial_\mu{}\,^*R^{\mu\nu}=0.
\end{equation}
In NLE however the two tensors $R_{\mu\nu}$ and $^*R_{\mu\nu}$ remain independent of one another, while it is only the former that is restricted by the NLE
equations of motion through $\partial_\mu R^{\mu\nu}=0$.

In order to exploit the potential offered by the use of the RS vector $\vec{R}$ let us effect the following change of variables for the fields $(\vec{D},\vec{B})$,
\begin{equation}
\vec{R}=\vec{D}+i\vec{B},\quad \vec{R}^*=\vec{D}-i\vec{B}\quad \Longleftrightarrow\quad
\vec{D}=\frac{1}{2}\left(\vec{R}+\vec{R}^*\right),\quad
\vec{B}=-\frac{1}{2}i\left(\vec{R}-\vec{R}^*\right),
\end{equation}
leading to ${\cal H}(\vec{D},\vec{B})={\cal H}_{RS}(\vec{R},\vec{R}^*)$ as well as
\begin{equation}
\vec{S}(\vec{R},\vec{R}^*)=-2i\frac{\partial{\cal H}_{RS}(\vec{R},\vec{R}^*)}{\partial\vec{R}^*},\qquad
\vec{S}^*(\vec{R},\vec{R}^*)=2i\frac{\partial{\cal H}_{RS}(\vec{R},\vec{R}^*)}{\partial\vec{R}},
\label{eq:S-RS}
\end{equation}
where $\vec{R}^*$ and $\vec{S}^*$ stand for the complex conjugates of $\vec{R}$ and $\vec{S}$.
Since $\vec{\nabla}\cdot\vec{R}=0$, the RS vector $\vec{R}$ derives from a complex
vector potential, $\vec{\Phi}(x^\mu)$, given in terms of the electric and magnetic vector potentials $(\vec{C},\vec{A})$
introduced previously for the fields $(\vec{D},\vec{B})$ and itself defined up to a gauge transformation by the gradient of an arbitrary complex scalar field, namely,
\begin{equation}
\vec{R}=\vec{D}+i\vec{B}=\vec{\nabla}\times\left(\vec{C} + i \vec{A}\right)=\vec{\nabla}\times\vec{\Phi},\quad
\vec{\Phi}=\vec{C} + i \vec{A},\quad
\vec{\Phi}'=\vec{\Phi}-\vec{\nabla}\varphi,\quad
\varphi=\varphi_e+i\varphi_m.
\end{equation}

In fact when full Poincar\'e invariance is in place this manifest 3d covariance for the 3-vector potentials $\vec{C}$ and $\vec{A}$ extends
to full Poincar\'e covariance in terms of the following 4-vectors $C^\mu$ and $A^\mu$ provided by all 3-scalar and 3-vector potentials involved
in the Hamiltonian formulation,
\begin{equation}
C^\mu=\left(\begin{array}{c}
C^0 \\ \vec{C} \end{array}\right),\qquad
A^\mu=\left(\begin{array}{c}
A^0 \\ \vec{A} \end{array}\right),
\end{equation}
with their associated field strengths,
\begin{equation}
F_{\mu\nu}=\partial_\mu A_\nu - \partial_\nu A_\mu,\qquad
C_{\mu\nu}=\partial_\mu C_\nu - \partial_\nu C_\mu,\qquad
^*C^{\mu\nu}=\frac{1}{2}\epsilon^{\mu\nu\rho\sigma}\,C_{\rho\sigma}.
\end{equation}
In particular,
\begin{equation}
C_{0i}=-H^i,\qquad
C_{ij}=-\epsilon^{ijk}\,D^k,\qquad
^*C_{0i}=D^i=G_{0i},,\qquad
^*C_{ij}=-\epsilon^{ijk}\,H^k=G_{ij},
\end{equation}
so that one identifies,
\begin{equation}
C_{\mu\nu}=-\,{}^*G_{\mu\nu},\qquad
^*C_{\mu\nu}=G_{\mu\nu},
\end{equation}
establishing at once the spacetime covariance of the 4-vector $C^\mu$ and its field strength $C_{\mu\nu}$.
The potential offered by the complex RS vector $\vec{R}$ may thus be extended to the following Poincar\'e covariant complex 4-vector potential and its field strength,
\begin{equation}
\Phi^\mu=C^\mu+iA^\mu,\qquad
\Phi_{\mu\nu}=\partial_\mu\Phi_\nu - \partial_\nu \Phi_\mu=C_{\mu\nu} + i F_{\mu\nu},\qquad
^*\Phi^{\mu\nu}=\frac{1}{2}\epsilon^{\mu\nu\rho\sigma}\Phi_{\rho\sigma},
\end{equation}
with its complex gauge transformations in the form
\begin{equation}
\Phi'_\mu=\Phi_\mu+\partial_\mu\varphi,\qquad
\Phi'_{\mu\nu}=\Phi_{\mu\nu},\qquad
\varphi=\varphi_e+i\varphi_m.
\end{equation}
In particular,
\begin{equation}
\Phi_{0i}=-S^i,\qquad
\Phi_{ij}=-\epsilon^{ijk}\,R^k,\qquad
^*\Phi_{0i}=R^i,\qquad
^*\Phi_{ij}=-\epsilon^{ijk}\,S^k.
\end{equation}
Therefore we simply obtain for the Riemann-Silberstein tensor,
\begin{equation}
R_{\mu\nu}={}^*C_{\mu\nu}+i {}^*F_{\mu\nu}={}^*\Phi_{\mu\nu},
\end{equation}
namely that $R_{\mu\nu}$ is  precisely the dual of the complex field strength $\Phi_{\mu\nu}$.
Furthermore the following identity may be established for that complex field strength and its dual,
\begin{equation}
\frac{1}{2}{}^*\Phi^{\mu\nu}\,\Phi^*_{\mu\nu}=
\frac{1}{4}\epsilon^{\mu\nu\rho\sigma}\,\Phi_{\mu\nu}\,\Phi^*_{\rho\sigma}=
\frac{1}{2}\Phi^{\mu\nu}\,\left({}^*\Phi^*_{\mu\nu}\right)=
\vec{S}\cdot\vec{R}^* + \vec{S}^*\cdot\vec{R}.
\label{eq:Ident1}
\end{equation}
Thus finally the fully Poincar\'e covariant NLE equations of motion $\partial_\mu R^{\mu\nu}=0$ are equivalent to
\begin{equation}
\partial_\mu{}^*\Phi^{\mu\nu}=0.
\end{equation}
In component form, with $\nu=0$ and $\nu=i$ ($i=1,2,3)$, respectively, of course one recovers the original NLE equations of motion,
\begin{equation}
\vec{\nabla}\cdot\vec{R}=0,\qquad
\vec{\nabla}\times\vec{S}=\partial_t\vec{R}.
\end{equation}
Beware that even though the equation $\partial_\mu{}^*\Phi^{\mu\nu}=0$ is indeed solved with $\Phi_{\mu\nu}=\partial_\mu\Phi_\nu-\partial_\nu\Phi_\mu$
(as being the Bianchi identity for the complex gauge field $\Phi^\mu$),
the complex 4-vector potential $\Phi_\mu$ which must be such that $\Phi_{0i}=-S^i$ still remains restricted by the constitutive relations which determine
the RS vector $\vec{S}=\vec{H}-i\vec{E}$ in terms of $\vec{\Phi}$ and of its relation to the Hamiltonian density ${\cal H}_{RS}(\vec{\nabla}\times\Phi,\vec{\nabla}\times\vec{\Phi}^*)$ --- while $\vec{\nabla}\times\vec{\Phi}=\vec{R}=\vec{D}+i\vec{B}$ ---, namely\footnote{In MLE this equation reduces simply to
$\vec{S}=-i\vec{R}=-i\vec{\nabla}\times\vec{\Phi}=\partial_t\vec{\Phi}+\vec{\nabla}\Phi^0$.},
\begin{equation}
-2i\frac{\partial{\cal H}_{RS}(\vec{R},\vec{R}^*)}{\partial\vec{R}^*}_{|_{\vec{R}=\vec{\nabla}\times\vec{\Phi}}}=\vec{S}(\vec{\nabla}\times\vec{\Phi},\vec{\nabla}\times\vec{\Phi}^*)=
\partial_t\vec{\Phi}+\vec{\nabla}\Phi^0,
\label{eq:Sol-gen}
\end{equation}
which is a gauge invariant condition. The NLE equations are thus completely solved in terms of a complex 4-vector gauge field
$\Phi^\mu$ provided the latter specific 3-vector nonlinear equation for its components be obeyed. All NLE nonlinearities solely reside in the RS 3-vector $\vec{S}$.
And as a consequence one has indeed $\vec{\nabla}\times\vec{S}=\partial_t\vec{R}$, and $\vec{\nabla}\cdot\vec{R}=0$.

Note that any given solution $\Phi^\mu$ to these equations is only defined up to an arbitrary gauge transformation, $\Phi'_\mu=\Phi_\mu+\partial_\mu\varphi$.
Hence in particular any solution may always be gauge transformed to the temporal axial gauge with ${\Phi^0}'=0$.

All of the above may be represented directly in terms of the first-order action (\ref{eq:S1-NLE}), upon the change of variables to the Riemann-Silberstein
complex fields. One then finds,
\begin{eqnarray}
\hspace{-10pt}S^{(1)}_{RS}[\Phi^\mu,{\Phi^\mu}^*\,]\!\!\! &=& \!\!\!\int_{(\infty)}d^4x^\mu\left\{
\frac{1}{4}i\partial_t\vec{\Phi}\cdot\left(\vec{\nabla}\times\vec{\Phi}^*\right) - \frac{1}{4}i\partial_t\vec{\Phi}^*\cdot\left(\vec{\nabla}\times\vec{\Phi}\right)
\,-\, \right. \nonumber \\
&& \left. \,-\,{\cal H}_{RS}(\vec{\nabla}\times\vec{\Phi},\vec{\nabla}\times\vec{\Phi}^*)\,+\,
\frac{1}{2}i\left[\vec{\nabla}\Phi^0\cdot\left(\vec{\nabla}\times\vec{\Phi}^*\right)-\vec{\nabla}{\Phi^0}^*\cdot\left(\vec{\nabla}\times\vec{\Phi}\right)\right]\right\}.
\end{eqnarray}
It readily follows that the gauge invariant Hamiltonian equations of motion for $\vec{\Phi}$ are recovered in the form,
\begin{equation}
\partial_t\vec{\Phi}=\vec{S}(\vec{\nabla}\times\vec{\Phi},\vec{\nabla}\times\vec{\Phi}^*) \,-\, \vec{\nabla}\Phi^0
=-2i\frac{\partial{\cal H}_{RS}(\vec{R},\vec{R}^*)}{\partial\vec{R}^*}_{|_{\vec{R}=\vec{\nabla}\times\vec{\Phi}}} \,-\,\vec{\nabla}\Phi^0.
\end{equation}

Given the first-order action $S^{(1)}_{RS}[\Phi^\mu,{\Phi^\mu}^*]$ above, it is now obvious that if the Hamiltonian density ${\cal H}_{RS}(\vec{R},\vec{R}^*)$ is
explicitly invariant under any global phase transformation of $\Phi^\mu$ into $e^{i\alpha}\Phi^\mu$ and thus of $\vec{R}$ into
$e^{i\alpha}\vec{R}$, this Hamiltonian action is left invariant. This implies that the NLE theory possesses this extra continuous U(1) or SO(2) global symmetry
which is the duality symmetry of ordinary electrodynamics but now extended to NLE. In view of the above definitions of the RS vector $\vec{S}$
in (\ref{eq:S-RS}) and of the RS tensor $R_{\mu\nu}$, clearly all these complex quantities transform with the same global phase factor,
\begin{equation}
{\Phi^\mu}'=e^{i\alpha}\,\Phi^\mu,\qquad
\Phi_{\mu\nu}'=e^{i\alpha}\,\Phi_{\mu\nu},\qquad
\vec{R}\,'=e^{i\alpha}\vec{R},\qquad
\vec{S}'=e^{i\alpha}\vec{S},\qquad
R'_{\mu\nu}=e^{i\alpha}R_{\mu\nu},
\end{equation}
which also translates into the fact that the two scalars or vectors of each of the pairs $(C^0,A^0)$, $(\vec{C},\vec{A})$, $(\vec{D},\vec{B})$
and $(\vec{H},-\vec{E})$ are rotated into one another by the same rotation angle $\alpha$.
Given the NLE equations of motion in the form of (\ref{eq:EOM-RS2}) or (\ref{eq:Sol-gen})
(which do not require Poincar\'e invariance as such), it is obvious that these are manifestly covariant as well under these duality
transformations. However the Lagrangian action of the NLE theory does not display in a manifest way that duality symmetry when it applies.

More specifically when ${\cal H}_{RS}(\vec{R},\vec{R}^*)$ is duality invariant we have
\begin{equation}
{\cal H}_{RS}(e^{i\alpha}\vec{R},e^{-i\alpha}\vec{R}^*)={\cal H}_{RS}(\vec{R},\vec{R}^*).
\end{equation}
By differentiation with respect to $\alpha$ at $\alpha=0$ this duality invariance translates into the identity\cite{Town,Gibbons},
\begin{equation}
\vec{R}\cdot\frac{\partial{\cal H}_{RS}(\vec{R},\vec{R}^*)}{\partial\vec{R}} \, - \,
\vec{R}^*\cdot\frac{\partial{\cal H}_{RS}(\vec{R},\vec{R}^*)}{\partial\vec{R}^*}=0,
\end{equation}
namely,
\begin{equation}
\vec{R}\cdot\vec{S}^* + \vec{R}^*\cdot\vec{S}=0,\qquad
\vec{H}\cdot\vec{D}-\vec{E}\cdot\vec{B}=0,
\label{eq:duality}
\end{equation}
or equivalently (see (\ref{eq:Ident1})),
\begin{equation}
\frac{1}{2}{}^*\Phi^{\mu\nu}\,\Phi^*_{\mu\nu}=\frac{1}{4}\epsilon^{\mu\nu\rho\sigma}\Phi_{\mu\nu}\,\Phi^*_{\rho\sigma}=
\frac{1}{2}\Phi^{\mu\nu}\,\left({}^*\Phi^*_{\mu\nu}\right)=0.
\end{equation}

Thus duality invariance is an invitation to consider the following two covariant 4-vector current densities, complex conjugates of one another,
\begin{equation}
J^\mu=\frac{1}{2}{}^*\Phi^{\mu\nu}\,\Phi^*_\nu=\frac{1}{4}\epsilon^{\mu\nu\rho\sigma}\,\Phi^*_\nu\,\Phi_{\rho\sigma},\qquad
{J^\mu}^*=\frac{1}{2}\left({}^*{\Phi^{\mu\nu}}^*\right)\,\Phi_\nu=\frac{1}{4}\epsilon^{\mu\nu\rho\sigma}\,\Phi_\nu\,\Phi^*_{\rho\sigma},\qquad
\end{equation}
which are conserved currents when duality invariance is in place, since,
\begin{equation}
\partial_\mu J^\mu=\frac{1}{4}{}^*\Phi^{\mu\nu}\,\Phi^*_{\mu\nu}=\partial_\mu{J^\mu}^*.
\end{equation}
In particular one finds for their time components,
\begin{eqnarray}
J^0 &=& \frac{1}{2}\vec{\Phi}^*\cdot\left(\vec{\nabla}\times\vec{\Phi}\right)=
\frac{1}{2}\left(\vec{C}\cdot\vec{D}+\vec{A}\cdot\vec{B}\right) + \frac{1}{2}i\left(\vec{C}\cdot\vec{B}-\vec{A}\cdot\vec{D}\right), \nonumber \\
{J^0}^* &=& \frac{1}{2}\vec{\Phi}\cdot\left(\vec{\nabla}\times\vec{\Phi}^*\right)=
\frac{1}{2}\left(\vec{C}\cdot\vec{D}+\vec{A}\cdot\vec{B}\right) - \frac{1}{2}i\left(\vec{C}\cdot\vec{B}-\vec{A}\cdot\vec{D}\right).
\end{eqnarray}
Let us then introduce the following electric, magnetic and mixed helicities (or vorticities) of the electromagnetic field
configuration, which measure the linking and knotting topological structure of the closed $\vec{D}$ and $\vec{B}$
field lines\cite{Review1,Review2}\footnote{The two mixed helicities $h_{em}$
and $h_{me}$ are identical on account of the identity
$\vec{V}\cdot(\vec{\nabla}\times\vec{W})=-\vec{\nabla}\cdot(\vec{V}\times\vec{W})+(\vec{\nabla}\times\vec{V})\cdot\vec{W}$,
and by discarding spacelike surface terms at infinity.},
\begin{equation}
h_e=\frac{1}{2}\int_{(\infty)}d^3\vec{x}\,\vec{C}\cdot\vec{D}=\frac{1}{2}\int_{\infty)}d^3\vec{x}\,\vec{C}\cdot\left(\vec{\nabla}\times\vec{C}\right),
\label{eq:he}
\end{equation}
\begin{equation}
h_m=\frac{1}{2}\int_{(\infty)}d^3\vec{x}\,\vec{A}\cdot\vec{B}=\frac{1}{2}\int_{\infty)}d^3\vec{x}\,\vec{A}\cdot\left(\vec{\nabla}\times\vec{A}\right),
\label{eq:hm}
\end{equation}
\begin{eqnarray}
h_{em} &=& \frac{1}{2}\int_{(\infty)}d^3\vec{x}\,\vec{C}\cdot\vec{B}=\frac{1}{2}\int_{(\infty)}d^3\vec{x}\,\vec{C}\cdot\left(\vec{\nabla}\times\vec{A}\right)= \nonumber \\
&=& \frac{1}{2}\int_{(\infty)}d^3\vec{x}\,\vec{A}\cdot\left(\vec{\nabla}\times\vec{C}\right)=\frac{1}{2}\int_{(\infty)}d^3\vec{x}\,\vec{A}\cdot\vec{D}=h_{me}.
\label{eq:hem}
\end{eqnarray}
Note that being given by the space integrations of the time components of the two conserved currents $J^\mu \pm {J^\mu}^*$, the combinations
$h_e+h_m$ and $h_{em}-h_{me}=0$ are in fact Lorentz scalars as well, when duality invariance is in place.
Furthermore the time evolution of these helicities is governed by the relations,
\begin{eqnarray}
\frac{d}{dt}h_e &=& \int_{(\infty)}d^3\vec{x}\,\vec{H}\cdot\vec{D},\qquad
\frac{d}{dt}h_m=-\int_{(\infty)}d^3\vec{x}\,\vec{E}\cdot\vec{B}, \nonumber \\
\frac{d}{dt}h_{em} &=& \frac{d}{dt}h_{me} = \frac{1}{2}\int_{(\infty)}d^3\vec{x}\left(\vec{H}\cdot\vec{B}-\vec{E}\cdot\vec{D}\right).
\label{eq:dt-h}
\end{eqnarray}
Consequently the scalar charges associated to the two currents $J^\mu$ and ${J^\mu}^*$ prove to be conserved,
identical and real, and are simply given by,
\begin{equation}
Q=\int_{(\infty)}d^3\vec{x}\,J^0=h_e + h_m =\int_{\infty)}d^3\vec{x}\,{J^0}^*=Q^*,
\end{equation}
while we have indeed explicitly $dQ/dt=dh_e/dt+dh_m/dt=0$ since, on account of duality invariance, $\vec{H}\cdot\vec{D}-\vec{E}\cdot\vec{B}=0$.
Thus in general the helicities $h_e$, $h_m$ and $h_{em}=h_{me}$ are not separately conserved quantities even when duality invariance of the NLE theory
is in place which implies the conservation only of $Q$.
On the other hand, let us note that even though the definitions of the charge $Q$ and the helicities $h_e$, $h_m$ and $h_{em}=h_{me}$
explicitly involve the gauge dependent vector potentials $\vec{\Phi}$, $\vec{C}$ and $\vec{A}$, each of these charges is a gauge invariant quantity
nonetheless since $\vec{\nabla}\cdot\vec{D}=0=\vec{\nabla}\cdot\vec{B}$ (and by discarding a spacelike surface term at infinity).

As a matter of fact the charge $Q$ is precisely also the conserved Noether charge associated to the U(1) duality symmetry of the NLE theory and
of its Hamiltonian action $S^{(1)}_{RS}[\Phi^\mu,{\Phi^\mu}^*]$, and the generator of that symmetry\cite{Deser1,Deser2}.
On account of Noether's first theorem\cite{Gov} it may readily
be established that the corresponding conserved spacetime covariant Noether current density is simply given by
\begin{equation}
J^\mu_{\rm dual}=\frac{1}{2}\left(J^\mu+{J^\mu}^*\right),\qquad
\partial_\mu J^\mu_{\rm dual}=0,
\end{equation}
with in particular thus the following Noether charge,
\begin{eqnarray}
Q_{\rm dual} &=& \frac{1}{4}\int_{(\infty)}d^3\vec{x}
\left[\vec{\Phi}\cdot\left(\vec{\nabla}\times\vec{\Phi}^*\right) + \vec{\Phi}^*\cdot\left(\vec{\nabla}\times\vec{\Phi}\right)\right]= \nonumber \\
&=& \frac{1}{4}\int_{(\infty)}d^3\vec{x}\left[\vec{\Phi}\cdot\vec{R}^*+\vec{\Phi}^*\cdot\vec{R}\right]=\frac{1}{2}\left(Q+Q^*\right)=Q=h_e+h_m.
\end{eqnarray}

Finally when Poincar\'e invariance is in place, the associated spacetime symmetry transformations are generated as well
by a collection of conserved Noether charges, which for spacetime translations correspond to the total energy, $E$, and momentum, $\vec{P}$,
of any field configuration obeying the NLE equations of motion, with,
\begin{equation}
E=\int_{(\infty)}d^3\vec{x}\,{\cal H}_{RS}(\vec{R},\vec{R}^*)=\int_{(\infty)}d^3\vec{x}\,{\cal H}(\vec{D},\vec{B}),\quad
\vec{P}=\int_{(\infty)}d^3\vec{x}\,\vec{D}\times\vec{B}=\frac{1}{2}i\,\int_{(\infty)}d^3\vec{x}\,\vec{R}\times\vec{R}^*,
\end{equation}
thereby also determining its relativistic invariant mass,
\begin{equation}
M=\sqrt{E^2-\vec{P}^2}.
\end{equation}

\subsection{Constructing solutions and generalised Bateman potentials}
\label{Sect2.3}

Given an arbitrary NLE theory within the first-order formalism and with an Hamiltonian density ${\cal H}_{NLE}(\vec{D},\vec{B})={\cal H}_{RS}(\vec{R},\vec{R}^*)$
let us consider its equations of motion,
\begin{equation}
\vec{\nabla}\cdot\vec{D}=0,\qquad \vec{\nabla}\times\vec{H}-\partial_t\vec{D}=\vec{0},\qquad
\vec{\nabla}\cdot\vec{B}=0,\qquad \vec{\nabla}\times\vec{E}+\partial_t\vec{B}=\vec{0},
\end{equation}
where
\begin{equation}
\vec{E}(\vec{D},\vec{B})=\frac{\partial{\cal H}_{NLS}(\vec{D},\vec{B})}{\partial\vec{D}},\qquad
\vec{H}(\vec{D},\vec{B})=\frac{\partial{\cal H}_{NLE}(\vec{D},\vec{B})}{\partial\vec{B}},
\end{equation}
or equivalently in complex RS form,
\begin{equation}
\vec{\nabla}\cdot\vec{R}=0,\qquad \vec{\nabla}\times\vec{S}=\partial_t\vec{R},
\end{equation}
where
\begin{equation}
\vec{S}(\vec{R},\vec{R}^*)=-2i\frac{\partial{\cal H}_{RS}(\vec{R},\vec{R}^*)}{\partial\vec{R}^*},
\end{equation}
or still equivalently, in terms of the complex 4-vector gauge potential $\Phi^\mu$,
\begin{equation}
\vec{R}=\vec{\nabla}\times\vec{\Phi},\qquad
-2i\frac{\partial{\cal H}_{RS}(\vec{R},\vec{R}^*)}{\partial\vec{R}^*}_{|_{\vec{R}=\vec{\nabla}\times\vec{\Phi}}}=\vec{S}(\vec{\nabla}\times\vec{\Phi},\vec{\nabla}\times\vec{\Phi}^*)=
\partial_t\vec{\Phi}+\vec{\nabla}\Phi^0.
\end{equation}

An obvious and trivial class of solutions is given by arbitrary static and homogeneous (spacetime independent)
fields $\vec{D}_0$ and $\vec{B}_0$, or $\vec{R}_0=\vec{D}_0+i\vec{B}_0$,
and the corresponding values for $\vec{E}_0(\vec{D}_0,\vec{B}_0)$, $\vec{H}_0(\vec{D}_0,\vec{B}_0)$ and
$\vec{S}_0(\vec{R}_0,\vec{R}^*_0)=\vec{H}_0-i\vec{E}_0$. Up to arbitrary gauge transformations a choice of 4-vector potential associated to
such a configuration is given by,
\begin{equation}
\Phi^0(t,\vec{x})=0,\qquad
\vec{\Phi}(t,\vec{x})=\frac{1}{2}\,\vec{R}_0\times\vec{x}\,+\,\vec{S}_0(\vec{R}_0,\vec{R}^*_0)\,t.
\end{equation}
This trivial remark is nonetheless relevant towards the construction of deformed HR knots 
in ModMax theories hereafter\footnote{Even though such static and homogeneous fields are {\sl bona fide} mathematical solutions to the differential equations,
they do not qualify as physical ones since their energy and momentum, for instance, are not finite.}.

More generally when spacetime symmetries are in place new solutions may be obtained by applying such spacetime transformations to already known
solutions. In the case of Poincar\'e invariance one may argue that in fact it is a same and physically unique field configuration that is being transformed
into itself but being observed from transformed inertial reference frames. However in the case of spacetime conformal symmetries
which may extend Poincar\'e invariance, such conformal transformations produce genuine physically distinct new solutions from known ones.
This feature is also put to use in the construction of deformed HR knots in ModMax theories.

In actual fact when working in terms of the complex RS representation of fields, spacetime translations may also be used to generate physically
distinct new solutions, by promoting the spacetime translation constant parameters to complex values\cite{Review1,Review2}.
If a specific field configuration $(\vec{R}(x^\mu),\vec{S}(x^\mu))$ or $\Phi^\mu(x^\mu)$ is a solution to the NLE equations, obviously so remains
the spacetime translated configuration $(\vec{R}(x^\mu+a^\mu),\vec{S}(x^\mu+a^\mu))$ or $\Phi^\mu(x^\mu+a^\mu)$ where $a^\mu$ is
an arbitrary constant spacetime 4-vector. Clearly even when the parameters $a^\mu$
are now promoted to take any constant complex values, this new configuration still defines a solution to the NLE equations. However this new
solution now describes a physically distinct one for the real and imaginary components of the complex RS vectors
$(\vec{R}(x^\mu+a^\mu),\vec{S}(x^\mu+a^\mu))$ and the complex gauge field $\Phi^\mu(x^\mu+a^\mu)$,
corresponding to new and physically distinct configurations for the fields
$(\vec{D}(x^\mu),\vec{B}(x^\mu),\vec{E}(x^\mu),\vec{H}(x^\mu))$, and this independently
of the nonlinearities inherent to the Hamiltonian density ${\cal H}_{RS}(\vec{R},\vec{R}^*)$. Such a complex constant spacetime translation
is also involved in the construction of deformed HR knots in ModMax theories.

In the specific case of MLE with $\vec{S}=-i\vec{R}=-i(\vec{E}+i\vec{B})$, Bateman designed a general method for the construction of solutions to
source-free Maxwell electrodynamics in vacuum\cite{Review1,Review2}, in terms of two complex Bateman scalar potentials $\alpha_0(x^\mu)$ and $\beta_0(x^\mu)$
obeying the self-dual property\footnote{In spacetime covariant form the self-duality property is expressed as
$(\partial_\mu\alpha_0 \partial_\nu\beta_0-\partial_\nu\alpha_0 \partial_\mu\beta_0)=i\epsilon_{\mu\nu\rho\sigma}\partial^\rho\alpha_0\partial^\sigma\beta_0$,
while the MLE solution is then given by $R_{\mu\nu}=-\epsilon_{\mu\nu\rho\sigma} \partial^\rho\alpha_0 \partial^\sigma\beta_0=
i(\partial_\mu\alpha_0\partial_\nu\beta_0 - \partial_\nu\alpha_0 \partial_\mu\beta_0)$.}
\begin{equation}
\vec{\nabla}\alpha_0\times\vec{\nabla}\beta_0=i\left(\partial_t\alpha_0\,\vec{\nabla}\beta_0\,-\,\partial_t\beta_0\,\vec{\nabla}\alpha_0\right).
\end{equation}
Indeed the MLE equations are then solved with $\vec{R}=\vec{\nabla}\alpha_0\times\vec{\nabla}\beta_0$ and
$\vec{S}=\partial_t\alpha_0\,\vec{\nabla}\beta_0\,-\,\partial_t\beta_0\,\vec{\nabla}\alpha_0$, since in that case
${\cal H}_{RS}(\vec{R},\vec{R}^*)=\vec{R}\cdot\vec{R}^*/2$ so that $\vec{S}=-i\vec{R}$. These Bateman solutions are necessarily such that
$\vec{R}^2=-i\vec{R}\cdot\vec{S}=0$, hence describing null $(\vec{E},\vec{B})$ fields with $\vec{E}^2-\vec{B}^2=0=\vec{E}\cdot\vec{B}$.

In order to extend the Bateman approach, and this within the context of NLE as well, let us begin by considering two complex scalar
fields\footnote{These are indeed scalar fields under the full Poincar\'e group if Poincar\'e invariance is in place for the NLE theory.} $\alpha(x^\mu)$
and $\beta(x^\mu)$, namely would-be Bateman potentials, which solve the first NLE equation of motion $\vec{\nabla}\cdot\vec{R}=0$ in the form,
\begin{equation}
\vec{R}=\vec{\nabla}\alpha\times\vec{\nabla}\beta\,-\,\vec{\nabla}\times\vec{\sigma},
\end{equation}
where for the sake of the present argument we have accounted for the possibility of an extra contribution to this RS vector in the form of the curl
of yet another complex 3-vector field $\vec{\sigma}$. Note that we then have,
\begin{equation}
\partial_t\vec{R}=\vec{\nabla}\times\left(\partial_t\alpha\, \vec{\nabla}\beta - \partial_t\beta\, \vec{\nabla}\alpha - \partial_t\vec{\sigma}\right),\qquad
\vec{R}=\vec{\nabla}\times\left(\lambda \alpha\vec{\nabla}\beta - (1-\lambda) \beta\vec{\nabla}\alpha\,-\,\vec{\sigma}\right),
\label{eq:R-curl}
\end{equation}
where $\lambda$ is an arbitrary complex constant --- in fact a gauge transformation parameter --- for which canonical values
are $\lambda=0,1/2,1$.
Therefore the NLE equation of motion involving the second RS vector reads,
\begin{equation}
\vec{\nabla}\times\left(\vec{S}-\left(\partial_t\alpha \vec{\nabla}\beta - \partial_t\beta \vec{\nabla}\alpha\right) + \partial_t\vec{\sigma}\right)=\vec{0}.
\end{equation}
Consequently the potentials $(\alpha,\beta)$ solve the NLS equations of motion provided they are such that,
\begin{equation}
-2i\frac{\partial{\cal H}_{RS}(\vec{R},\vec{R}^*)}{\partial\vec{R}^*}_{|_{\vec{R}=\vec{\nabla}\alpha\times\vec{\nabla}\beta - \vec{\nabla}\times\vec{\sigma}}}\,-\,
\left(\partial_t\alpha \vec{\nabla}\beta - \partial_t\beta \vec{\nabla}\alpha\right) + \partial_t\vec{\sigma} =\vec{\nabla}\sigma^0,
\label{eq:Bateman-potentials}
\end{equation}
namely such that the l.h.s.~of this expression reduces to the gradient of yet some other 3-scalar field $\sigma^0(x^\mu)$ which is implicitly defined through this identity.

Hence as a matter of fact, the two fields $\sigma^0$ and $\vec{\sigma}$ combine into a covariant 4-vector $\sigma^\mu=(\sigma^0,\vec{\sigma})$, while if they
obey the above remaining equation of motion (\ref{eq:Bateman-potentials}) such a solution corresponds, up to an arbitrary gauge transformation,
to a 4-vector gauge potential given in the spacetime covariant form,
\begin{equation}
\Phi_\mu=-\lambda\,\alpha\,\partial_\mu \beta + (1-\lambda)\, \beta\,\partial_\mu\alpha\,+\,\sigma_\mu,
\end{equation}
namely,
\begin{equation}
\Phi^0 = -\lambda\,\alpha\,\partial_t\beta\,+\,(1-\lambda)\,\beta\,\partial_t\alpha\,+\,\sigma^0, \qquad
\vec{\Phi} = \lambda\,\alpha\,\vec{\nabla}\beta\,-\,(1-\lambda)\,\beta\,\vec{\nabla}\alpha\,-\,\vec{\sigma}.
\end{equation}
Correspondingly, the associated electric and magnetic vector potentials $\vec{C}$ and $\vec{A}$ are readily expressed in terms of the Bateman
potentials as follows,
\begin{equation}
\vec{\Phi}=\vec{C}+i\vec{A}=\alpha\vec{\nabla}\beta-\vec{\nabla}\left((1-\lambda)\alpha\beta\right)-\vec{\sigma}
=\lambda\alpha\vec{\nabla}\beta - (1-\lambda)\beta\vec{\nabla}\alpha - \vec{\sigma}.
\end{equation}
Note that the 4-vector potential $\sigma^\mu$ is only defined up to a gauge transformation, $\sigma'_\mu=\sigma_\mu+\partial_\mu\varphi$. 

The extended set of Bateman potentials $(\alpha,\beta,\sigma^\mu)$ thus determines a solution to the NLE equations provided these potentials be such that
\begin{equation}
\vec{S}=-2i\frac{\partial{\cal H}_{RS}(\vec{R},\vec{R}^*)}{\partial\vec{R}^*}_{|_{\vec{R}=\vec{\nabla}\alpha\times\vec{\nabla}\beta - \vec{\nabla}\times\vec{\sigma}}}=
\partial_t\vec{\Phi}+\vec{\nabla}\Phi^0=\left(\partial_t\alpha\, \vec{\nabla}\beta - \partial_t\beta\, \vec{\nabla}\alpha\right) - \partial_t\vec{\sigma}
+ \vec{\nabla}\sigma^0.
\label{eq:Bateman-potentials2}
\end{equation}
When $\sigma^\mu\ne 0$ we shall refer to the set $(\alpha,\beta,\sigma^\mu)$ and such a solution as a {\it generalised} Bateman solution with its
{\it generalised} Bateman potentials, while if $\sigma^\mu=0$ simply as a Bateman solution with its usual Bateman potentials $(\alpha,\beta)$.

Let us point out that one has
\begin{equation}
\vec{R}\cdot\vec{S}=\left(\vec{D}+i\vec{B}\right)\cdot\left(\vec{H}-i\vec{E}\right)=\left(\vec{H}\cdot\vec{D}+\vec{E}\cdot\vec{B}\right)
+i\left(\vec{H}\cdot\vec{B}-\vec{E}\cdot\vec{D}\right),
\end{equation}
while
\begin{equation}
\vec{R}\cdot\vec{S}_{|_{\vec{R}=\vec{\nabla}\alpha\times\vec{\nabla}\beta -\vec{\nabla}\times\vec{\sigma}}}=
\left(\vec{\nabla}\alpha\times\vec{\nabla}\beta\right)\cdot\left(-\partial_t\vec{\sigma}+\vec{\nabla}\sigma^0\right)
-\left(\vec{\nabla}\times\vec{\sigma}\right)\cdot\left(\partial_t\alpha\, \vec{\nabla}\beta - \partial_t\beta\, \vec{\nabla}\alpha - \partial_t\vec{\sigma}
+ \vec{\nabla}\sigma^0\right).
\end{equation}
Therefore given a Bateman solution a nonvanishing value for $\vec{R}\cdot\vec{S}$ signals the presence of the terms in
$\sigma^\mu$ in the above equation (\ref{eq:Bateman-potentials2})
for the Bateman potentials for some field $\sigma^\mu(x^\mu)$. On the other hand even if $\vec{R}\cdot\vec{S}=0$
and $\sigma^\mu=0$, this does certainly not imply that the fields $(\vec{D},\vec{B})$, $(\vec{E},\vec{H})$ or $(\vec{E},\vec{B})$
produced by such a Bateman solution are necessarily null fields, in contradistinction to the Maxwell theory case. Note that the original Bateman
construction implicitly assumes from the outset that no field $\sigma^\mu$ is included with the Bateman potentials,
as is implied by their assumed self-duality property in the MLE case. Both situations may occur however, when looking for more general
solutions, whether in MLE or in NLE such as ModMax theories.

A final remark may be in order. If a field configuration which solves the NLE equations may be represented in terms of generalised
Bateman potentials $(\alpha,\beta,\sigma^\mu)$, then this representation is certainly not unique. Besides gauge transformations of $\sigma^\mu$,
obviously such potentials are ever defined only up to arbitrary complex constants, a transformation which
simply generates as well a local gauge transformation of $\Phi^\mu$. In addition if only for such transformations, the two potentials $(\alpha,\beta)$ may be permuted
possibly together with an alternating change of sign, $(\alpha\rightarrow \pm\beta, \beta\rightarrow\mp \alpha))$, and then
as well jointly with $\lambda\rightarrow (1-\lambda)$; or else they may be rescaled by arbitrary complex scale factor such
that $(\alpha,\beta)\rightarrow(s\alpha,\beta/s)$ with $s\in\mathbb{C}$, without a modification of the gauge potential $\Phi^\mu$.
On the other hand one should expect that there exist solutions to the NLE equations which do not possess a Bateman representation, even when allowing
for the nonvanishing contribution in $\sigma^\mu$ to the necessary condition (\ref{eq:Bateman-potentials2}) that the generalised
Bateman potentials ought to obey.

\section{ModMax Nonlinear Electrodynamics}
\label{Sect3}

From here on let us restrict specifically to ModMax theories\cite{Town}, which are one-parameter continuous deformations of ordinary linear
electrodynamics in terms of the single positive parameter $\gamma\ge 0$. Their action is given in terms of the
Lagrangian density\cite{Town,Kosy}
\begin{eqnarray}
{\cal L}_\gamma({\cal S},{\cal P}) &=& \cosh\gamma\,{\cal S}\,+\,\sinh\gamma\,\sqrt{{\cal S}^2+{\cal P}^2}, \nonumber \\
{\cal L}_\gamma(\vec{E},\vec{B})&=& \frac{1}{2}\cosh\gamma\left(\vec{E}^2-\vec{B}^2\right)+\frac{1}{2}\sinh\gamma\,
\sqrt{\left(\vec{E}^2-\vec{B}^2\right)^2+4\left(\vec{E}\cdot\vec{B}\right)^2},
\end{eqnarray}
while in first-order form their Hamiltonian density is expressed as\cite{Town}
\begin{eqnarray}
{\cal H}_\gamma(\vec{D},\vec{B}) &=& \frac{1}{2}\cosh\gamma\left(\vec{D}^2+\vec{B}^2\right)-\frac{1}{2}\sinh\gamma
\sqrt{\left(\vec{D}^2-\vec{B}^2\right)^2+4\left(\vec{D}\cdot\vec{B}\right)^2}, \nonumber \\
{\cal H}^{(RS)}_{\gamma}(\vec{R},\vec{R}^*) &=& \frac{1}{2}\cosh\gamma\,\vec{R}\cdot\vec{R}^* - \frac{1}{2}\sinh\gamma\sqrt{\vec{R}^2\,\vec{R}^{*2}},
\end{eqnarray}
with thus in particular,
\begin{equation}
\vec{S}(\vec{R},\vec{R}^*)=-i\cosh\gamma\,\vec{R}\,+\,i\sinh\gamma\frac{\vec{R}^2}{\sqrt{\vec{R}^2\,\vec{R}^{*2}}}\,\vec{R}^*.
\label{eq:S-ModMax}
\end{equation}
As established in Ref.\cite{Town} these ModMax theories are the only NLE theories which preserve all the same symmetries as those of Maxwell linear
electrodynamics (recovered for $\gamma=0$) --- namely spacetime Poincar\'e and conformal invariance as is manifest from their Lagrangian density, as well as
duality invariance as is manifest from their Hamiltonian density --- and  which are continuous deformations of MLE. The only other such
electrodynamics theory sharing all these same symmetries is defined\cite{Town} by the BB Hamiltonian density\cite{BB}, ${\cal H}_{BB}=|\vec{D}\times\vec{B}|$.
Note that ${\cal L}_\gamma$ is nonanalytic in $({\cal S},{\cal P})$ at $({\cal S},{\cal P})=(0,0)$ for all $\gamma\ne 0$, namely for null $(\vec{E},\vec{B})$ fields.

In Ref.\cite{Town} a solution to this dynamics has explicitly been constructed in the form of an elliptically polarised monochromatic travelling
plane wave, which in the linear limit $\gamma=0$ reduces to a circularly polarised null plane wave.
Before finally turning to deformed HR knots in the next section, here the Bateman approach is briefly illustrated with two simple classes of solutions
in the next two Subsections.

But let us still point out the following general property however, valid for any Bateman solution such that $\sigma^\mu=0$
in (\ref{eq:Bateman-potentials2}). This property extends to the very specific case of ModMax theories a same property valid for the ordinary Bateman solutions
to the linear Maxwell equations in the form recalled above\cite{Review1,Review2,BB2}. Namely that given any pair $(\alpha(x^\mu),\beta(x^\mu))$
of Bateman potentials obeying (\ref{eq:Bateman-potentials2}) but specifically with $\sigma^\mu(x^\mu)=0$, then any bi-holomorphic complex
transformation $(f(\alpha,\beta),g(\alpha,\beta))$ defines
again a pair of Bateman potentials solving once again the ModMax equations in (\ref{eq:Bateman-potentials2}) and once again with
$\sigma^\mu=0$. This simple observation provides a powerful tool to generate whole classes of new solutions to the ModMax equations starting
from known ones, provided of course that no term in $\sigma^\mu$ be present for the initial solution. The basic simple double reason for this fact is,
on the one hand, the following property,
\begin{equation}
\partial_\mu f\,\partial_\nu g - \partial_\nu\, f \partial_\mu g=\left(\partial_\alpha f \partial_\beta g - \partial_\beta f \partial_\alpha g\right)
\left(\partial_\mu\alpha\, \partial_\nu\beta - \partial_\nu\alpha\, \partial_\mu \beta\right),
\end{equation}
namely in 3-vector form,
\begin{eqnarray}
\vec{R}(f,g) = \vec{\nabla}f\times\vec{\nabla}g &=&
\left(\partial_\alpha f \partial_\beta g - \partial_\beta f \partial_\alpha g\right)\left(\vec{\nabla}\alpha\times\vec{\nabla}\beta\right)=
\left(\partial_\alpha f \partial_\beta g - \partial_\beta f \partial_\alpha g\right)\vec{R}(\alpha,\beta), \nonumber \\
\partial_t f \,\vec{\nabla}g - \partial_t g\,\vec{\nabla}f &=& \left(\partial_\alpha f \partial_\beta g - \partial_\beta f \partial_\alpha g\right)
\left(\partial_t\alpha\,\vec{\nabla}\beta - \partial_t\beta\,\vec{\nabla}\alpha\right),
\end{eqnarray}
as well as, on the other hand, the specific structure in $\vec{R}$ and $\vec{R}^*$ for the second term on the r.h.s.~of (\ref{eq:S-ModMax})
expressing the RS vector $\vec{S}$ for ModMax theories, implying that likewise in this very specific case, in a self-explanatory notation,
\begin{equation}
\vec{S}(f,g)=\left(\partial_\alpha f \partial_\beta g - \partial_\beta f \partial_\alpha g\right)\, \vec{S}(\alpha,\beta).
\end{equation}
In other words in the specific case of ModMax theories their equations of motion are covariant under such bi-holomorphic transformations
of the scalar potentials of ordinary Bateman solutions.

\subsection{Static and Homogeneous Fields}
\label{Sect3.1}

Any spacetime constant field configuration $(\vec{D}_0,\vec{B}_0)$ is a trivial solution to the ModMax equations,
with the associated fields $(\vec{E}_0,\vec{H}_0)$ taking values such that,
\begin{equation}
\vec{H}_0 - i \vec{E}_0= \vec{S}_0=-i\cosh\gamma\,\vec{R}_0\,+\,i\sinh\gamma\frac{\vec{R}^2_0}{\sqrt{\vec{R}^2_0\,\vec{R}^{*2}_0}}\,\vec{R}^*_0,\qquad
\vec{R}_0=\vec{D}_0+i\vec{B}_0,
\end{equation}
and thus provided that the vectors $(\vec{D}_0,\vec{B}_0)$ are not null, $\vec{R}^2_0\ne 0$ (namely such that at least one of the following two conditions
is met, $\vec{D}^2_0\ne \vec{B}^2_0$ or $\vec{D}_0\cdot\vec{B}_0\ne 0$).
Under which circumstances may generalised Bateman potentials be identified to represent such a solution?

Consider first the general situation when the two vectors $\vec{D}_0$ and $\vec{B}_0$ are not aligned, namely
\begin{equation}
\vec{D}_0\times\vec{B}_0\ne \vec{0},
\end{equation}
so that they generate a plane. Let us use a parameter $\eta=\pm 1$ to distinguish the orientation of that plane, and then introduce a unit vector normal
to that plane defined by
\begin{equation}
\hat{u}_3=\eta\,\frac{\vec{D}_0\times\vec{B}_0}{|\vec{D}_0\times\vec{B}_0|},\qquad \eta=\pm 1,
\end{equation}
in order that the triad $(\eta\vec{D}_0,\vec{B}_0,\hat{u}_3)$, in that order, is a right-handed frame.

A class of generalised Bateman potentials which reproduces this field configuration is then found in the following form, the potential $\sigma^\mu$
remaining to be identified,
\begin{eqnarray}
\alpha(t,\vec{x}) &=& \left(\hat{u}_3+s(\vec{R}_0\times\hat{u}_3)\right)\cdot\vec{x} + \alpha_0\,|\vec{D}_0\times\vec{B}_0|\, t + \alpha_c, \nonumber \\
\beta(t,\vec{x}) &=& \left(\vec{R}_0\times\hat{u}_3\right)\cdot\vec{x} + \beta_0\,|\vec{D}_0\times\vec{B}_0|\, t + \beta_c,
\end{eqnarray}
where $s$, $\alpha_0$, $\beta_0$, $\alpha_c$ and $\beta_c$ are arbitrary complex constants. Given the equation of motion (\ref{eq:Bateman-potentials2})
the 4-vector potential $\sigma^\mu$ is restricted to be such that,
\begin{equation}
\vec{\nabla}\times\vec{\sigma}=\vec{\nabla}\alpha\times\vec{\nabla}\beta-\vec{R}_0=\vec{0},\qquad
\vec{\nabla}\sigma^0-\partial_t\vec{\sigma}=\vec{S}_0-\left(\partial_t\alpha\,\vec{\nabla}\beta - \partial_t\beta\,\vec{\nabla}\alpha\right)=\vec{\Sigma}_0,
\end{equation}
where the vector $\vec{\Sigma}_0$ denotes the following quantity, decomposed in the frame $(\eta\vec{D}_0,\vec{B}_0,\hat{u}_3)$,
\begin{eqnarray}
\vec{\Sigma}_0 &=& \left[-i\left(\cosh\gamma - \sinh\gamma\frac{\vec{R}^2_0}{\sqrt{\vec{R}^2_0\vec{R}^{*2}_0}}\right)\,-\,
\eta(\alpha_0-s\beta_0)\vec{R}_0\cdot\vec{B}_0\right]\,\vec{D}_0 \,+   \\
&&\hspace{-10pt} +\left[\left(\cosh\gamma + \sinh\gamma\frac{\vec{R}^2_0}{\sqrt{\vec{R}^2_0\vec{R}^{*2}_0}}\right)\,+\,
\eta(\alpha_0 - s\beta_0) \vec{R}_0\cdot\vec{D}_0\right]\,\vec{B}_0\,+\,\eta\beta_0\,\vec{D}_0\times\vec{B}_0 \nonumber
\end{eqnarray}
(note how the parameters $s$, $\alpha_0$ and $\beta_0$ contribute through the sole combination $(\alpha_0 - s\beta_0)$).
Therefore a general choice for the potential $\sigma^\mu(x^\mu)$ is such that,
\begin{equation}
\sigma^0(t,\vec{x})=\vec{x}\cdot\vec{\Sigma}_0\,-\,\partial_t\varphi_0(t,\vec{x})+\chi_0(t),\qquad
\vec{\sigma}(t,\vec{x})=-\vec{\nabla}\varphi_0(t,\vec{x}),
\end{equation}
where $\varphi_0(t,\vec{x})$ and $\chi_0(t)$ arbitrary complex functions. However upon the gauge transformation with
\begin{equation}
\varphi(t,\vec{x})=\varphi_0(t,\vec{x})-\tilde{\chi}_0(t),\qquad
\frac{d\tilde{\chi}_0(t)}{dt}=\chi_0(t),
\end{equation}
this solution is gauge equivalent to the final choice of 4-vector Bateman potential $\sigma^\mu(x^\mu)$,
\begin{equation}
\sigma^0(t,\vec{x})=\vec{x}\cdot\vec{\Sigma}_0,\qquad
\vec{\sigma}(t,\vec{x})=\vec{0}.
\end{equation}
Thus quite generally, unless the field configuration is restricted by specific additional conditions, the potential $\sigma^\mu(x^\mu)$ needs to be nonvanishing
for solutions generated through the generalised Bateman construction, even for as simple an electromagnetic field configuration as a spacetime constant one,
here such that $\vec{D}_0\times\vec{B}_0\ne\vec{0}$. Note that this observation applies as well even when $\gamma=0$ in the case of MLE.

A similar discussion applies when $\vec{D}_0\times\vec{B}_0=\vec{0}$, namely when both these fields are colinear to a common unit basis vector $\hat{u}_1$,
\begin{equation}
\vec{D}_0=D_0\,\hat{u}_1,\qquad \vec{B}_0=B_0\,\hat{u}_1,\qquad \vec{R}_0=R_0\,\hat{u}_1,\qquad R_0=D_0 + i B_0,
\end{equation}
where the real components $D_0$ and $B_0$ may be of either sign. Let us then extend the unit vector $\hat{u}_1$ to an orthonormalised right-handed frame
$(\hat{u}_1,\hat{u}_2,\hat{u}_3)$ with $\hat{u}_2$ and $\hat{u}_3$ perpendicular to each other and to $\hat{u}_1$. One then finds that the following
generalised Bateman potentials, once again up to a gauge transformation in $\sigma^\mu$,
\begin{equation}
\alpha(t,\vec{x})=\hat{u}_3\cdot\vec{x} + \alpha_0 t + \alpha_c,\quad
\beta(t,\vec{x}) = \left(\vec{R}_0\times\hat{u}_3\right)\cdot\vec{x} + \beta_0 t + \beta_c,\quad
\sigma^0(t,\vec{x})=\vec{x}\cdot\vec{\Xi}_0,\quad
\vec{\sigma}=\vec{0},
\end{equation}
where the constant vector $\vec{\Xi}_0$ is defined as,
\begin{equation}
\vec{\Xi}_0=-i\left(\cosh\gamma\,R_0\,-\,\sinh\gamma\,\frac{R^2_0}{\sqrt{R^2_0\,{R^2_0}^*}}\,R^*_0\right)\,\hat{u}_1\,+\,\alpha_0\,R_0\,\hat{u}_2\,+\,\beta_0\,\hat{u}_3,
\end{equation}
provide a generalised Bateman representation of such a field configuration. Hence once again a nonvanishing potential $\sigma^\mu$ is required in this case
as well, unless additional restrictions are imposed on the fields involved.

It is possible to restrict to Bateman potentials with $\sigma^\mu=0$, but then at the cost of restricting the set of spacetime constant
field solutions that possess a Bateman representation. For instance when $\vec{D}_0\times\vec{B}_0=\vec{0}$ and by requiring that $\sigma^0=0$,
one finds that necessarily all fields vanish identically, $\vec{D}_0=\vec{B}_0=\vec{E}_0=\vec{H}_0=\vec{0}$, irrespective of the value for $\gamma\ge 0$.
Likewise in the case when $\vec{D}_0\times\vec{B}_0\ne\vec{0}$ but restricting again to the condition $\sigma^0=0$, requires that $\beta_0=0$, with $s$ however,
remaining free to be chosen but no loss of generality is incurred by setting as well $s=0$. But more importantly the set of fields $(\vec{D}_0,\vec{B}_0)$
for which such Bateman potentials exist must meet the following condition for the value of the single parameter $\alpha_0$,
\begin{equation}
\alpha_0=-i\eta\frac{1}{\vec{R}_0\cdot\vec{B}_0}\left(\cosh\gamma - \sinh\gamma\frac{\vec{R}^2_0}{\sqrt{\vec{R}^2_0\vec{R}^{*2}_0}}\right)
=-\eta\frac{1}{\vec{R}_0\cdot\vec{D}_0}\left(\cosh\gamma + \sinh\gamma\frac{\vec{R}^2_0}{\sqrt{\vec{R}^2_0\vec{R}^{*2}_0}}\right)
=\alpha_0.
\end{equation}
Clearly this last condition (of which a detailed analysis is not developed here) determines a correlation between the values of $\vec{D}^2_0$, $\vec{B}^2_0$
and $\vec{D}_0\cdot\vec{B}_0$ as function of the ModMax parameter $\gamma\ge 0$. For instance for MLE with $\gamma=0$ this condition
requires $\vec{D}^2_0=\vec{B}^2_0$ (without necessarily $\vec{D}_0\cdot\vec{B}_0=0$).

As an explicit illustration with $\gamma\ne 0$ and $\sigma^\mu=0$,
let us restrict to the case that $\vec{D}_0\cdot\vec{B}_0=0$ with $\vec{B}_0\ne\vec{0}$, and thus as well
$\vec{D}^2_0\ne\vec{B}^2_0$. One then finds,
\begin{equation}
\frac{\vec{R}^2_0}{\sqrt{\vec{R}^2_0\vec{R}^{*2}_0}}=\frac{\vec{D}^2_0-\vec{B}^2_0}{|\vec{D}^2_0-\vec{B}^2_0|}
={\rm sign}\,(\vec{D}^2_0-\vec{B}^2_0)=\delta=\pm 1,
\end{equation}
together with the values,
\begin{equation}
|\vec{D}_0|=e^{\delta\gamma}\,|\vec{B}_0|,\qquad
\alpha_0=-\eta,\qquad \beta_0=0=s.
\end{equation}
In order to be explicit, let us align the right-handed orthonormalised cartesian frame $(\hat{e}_1,\hat{e}_2,\hat{e}_3)$ (related to the coordinate
system $\vec{x}=(x^1,x^2,x^3)=(x,y,z)$) with the right-handed triad $(\eta\vec{D}_0,\vec{B}_0,\hat{u}_3)$, so that
\begin{equation}
\vec{D}_0=\eta\,B_0\,e^{\delta\gamma}\,\hat{e}_1,\qquad
\vec{B}_0=B_0\,\hat{e}_2,\qquad B_0>0,\qquad \delta=\pm 1,\quad \eta=\pm 1,
\end{equation}
while,
\begin{equation}
\alpha(t,\vec{x})=x^3-\eta t + \alpha_c,\qquad
\beta(t,\vec{x})=iB_0(x^1+i\eta e^{\delta\gamma}\,x^2)+\beta_c,
\label{eq:Bateman-Constant}
\end{equation}
and thus finally,
\begin{equation}
\vec{H}_0-i\vec{E}_0=\vec{S}_0=\vec{S}(t,\vec{x})=\partial_t\alpha\vec{\nabla}\beta - \partial_t\beta \vec{\nabla}\alpha
=-i\eta B_0 \hat{e}_1 + B_0 e^{\delta\gamma}\hat{e}_2,
\end{equation}
namely,
\begin{equation}
\vec{E}_0=\eta\,B_0\,\hat{e}_1,\qquad
\vec{H}_0=B_0\,e^{\delta\gamma}\,\hat{e}_2.
\end{equation}
Even though the two pairs of fields $(\vec{E}_0,\vec{B}_0)$ and $(\vec{D}_0,\vec{H}_0)$ are thus null for whatever value of $\gamma\ge 0$,
the solution is well defined within the Hamiltonian formalism. This configuration will be of use in the construction of the ModMax deformed hopfion-Ran\~ada knots.

\subsection{Monochromatic Transverse Travelling Plane Wave Solutions}
\label{Sect3.2}

In a likewise manner it is possible to identify Bateman potentials associated to a monochromatic plane wave solution of the ModMax equations.
Consider such a travelling wave of wave vector $\vec{k}=k\hat{k}$ ($k\ge 0$) and angular frequency $\omega$, and $\hat{n}$ a unit vector perpendicular to $\vec{k}$.
A choice of Bateman potentials generating such a solution is given as
\begin{equation}
\alpha(t,\vec{x})=\alpha_0\,e^{-i(\omega t - \vec{k}\cdot\vec{x})},\qquad
\beta(t,\vec{x})=i\beta_1\,\left(\hat{n}\,+\,ie^{\delta\gamma}\,\hat{k}\times\hat{n}\right)\cdot\vec{x},\qquad \delta=\pm 1,
\end{equation}
where $\alpha_0,\beta_1\in\mathbb{C}$ are two arbitrary complex constants, while in order to solve the ModMax equations of motion,
and with a choice for the third Bateman potential such that $\sigma^\mu=0$, one also needs,
\begin{equation}
\omega=|\vec{k}\,|=k,\qquad \hat{n}\cdot\vec{k}=0.
\end{equation}
The corresponding plane waves thus propagate at the speed of light in vacuum, in spite of the nonlinearities inherent to ModMax dynamics.
For the associated electric and magnetic fields one finds, assuming here for simplicity that $\alpha_0\beta_1\in\mathbb{R}$ without loss of generality,
\begin{eqnarray}
\vec{D}(t,\vec{x}) &=&  k\,\alpha_0\beta_1\left[ e^{\delta\gamma}\sin(\omega t - \vec{k}\cdot\vec{x})\ \hat{n} \,-\,
\cos(\omega t - \vec{k}\cdot\vec{x})\ \hat{k}\times\hat{n}\right], \nonumber \\
\vec{B}(t,\vec{x}) &=&   k\,\alpha_0\beta_1\left[ e^{\delta\gamma}\cos(\omega t - \vec{k}\cdot\vec{x})\ \hat{n} \,+\,
\sin(\omega t - \vec{k}\cdot\vec{x})\ \hat{k}\times\hat{n}\right], \nonumber \\
\vec{E}(t,\vec{x}) &=&  k\,\alpha_0\beta_1\left[ \sin(\omega t - \vec{k}\cdot\vec{x})\ \hat{n} \,-\,
e^{\delta\gamma}\,\cos(\omega t - \vec{k}\cdot\vec{x})\ \hat{k}\times\hat{n}\right], \\
\vec{H}(t,\vec{x}) &=&  k\,\alpha_0\beta_1\left[ \cos(\omega t - \vec{k}\cdot\vec{x})\ \hat{n} \,+\,
e^{\delta\gamma}\,\sin(\omega t - \vec{k}\cdot\vec{x})\ \hat{k}\times\hat{n}\right]. \nonumber
\end{eqnarray}
Therefore these are elliptically polarised transverse travelling plane waves, becoming circularly polarised in the linear limit $\gamma=0$
(when $\alpha_0 \beta_1$ is complex the principal axes of the elliptic polarisation are simply rotated in the plane perpendicular to $\hat{k}$).

Direct calculations also establish that the pairs $(\vec{E},\vec{B})$ and $(\vec{D},\vec{H})$ are null fields throughout spacetime,
\begin{equation}
\vec{E}^2-\vec{B}^2=0=\vec{E}\cdot\vec{B},\qquad
\vec{D}^2-\vec{H}^2=0=\vec{D}\cdot\vec{H},
\end{equation}
while
\begin{equation}
{\cal H}_\gamma(\vec{D},\vec{B})=k^2\,\alpha_0^2\,\beta_1^2\,e^{\delta\gamma},\quad
\vec{D}\times\vec{B}=k^2\,\alpha_0^2\,\beta_1^2\,e^{\delta\gamma}\,\hat{k}=\vec{E}\times\vec{H},\quad
\vec{D}\cdot\vec{E}=k^2\,\alpha^2_0\,\beta^2_1\,e^{\delta\gamma}=\vec{H}\cdot\vec{B}.
\end{equation}
Even though these fields are {\sl bona fide} mathematical solutions to the ModMax equations, they do not qualify as physical ones since their energy and momentum,
for instance, are not finite. However when regularising the volume integral of the energy and momentum densities ${\cal H}_\gamma$
and $\vec{D}\times\vec{B}$ to a finite volume, one finds values for their total energy and momentum which are such that $\vec{P}=E\,\hat{k}$,
hence representing the propagation at the speed of light in vacuum
of an electromagnetic entity possessing a vanishing relativistic invariant mass.

\section{ModMax Deformed Hopfion-Ra\~nada Knots}
\label{Sect4}

Within the context of source-free Maxwell linear electrodynamics hopfion-Ra\~nada knots may be constructed from
a variety of approaches\cite{Review1,Review2}.
Besides the dual two Hopf maps for the electric and magnetic sectors\cite{HR1,HR2, Review1,Review2}, combinations of special conformal transformations
or conformal inversions
with pure imaginary time translations also produce these solutions starting from static and homogeneous null electric and magnetic fields.
Since ModMax theories share precisely the same global symmetries with MLE, the same transformations may be exploited to construct ModMax
deformed hopfion-Ra\~nada knots, which are continuous deformations in the ModMax parameter $\gamma\ge 0$ of the ordinary hopfion-Ra\~nada knots.
As established hereafter, to each hopfion-Ra\~nada knot of MLE there correspond two distinct deformations into ModMax hopfion-Ra\~nada knots
distinguished by a parameter $\delta=\pm 1$ taking only those two values (a situation on a par with that encountered above for the previous two
examples of ModMax solutions given in terms of Bateman potentials with $\sigma^\mu=0$). In the limit $\gamma=0$ these two ModMax knot solutions then
coalesce back into the corresponding ordinary hopfion-Ra\~nada knot of the source-free linear Maxwell theory.

As it turned out the present authors found it most efficient to apply the following steps and transformations to generate
the sought-for ModMax knot solutions, which in the case $\gamma=0$ also produce hopfion-Ra\~nada knots. First, within the Lagrangian formulation,
consider a configuration of static and homogeneous electric and magnetic fields, which is assumed to be non-null in order to avoid the singularities inherent
to the ModMax Lagrangian formulation for null fields. Apply then a conformal inversion to that configuration to identify a new configuration for new RS fields
$\vec{R}$ and $\vec{S}$ which obey the ModMax equations as well, on account of conformal covariance of the ModMax equations.
It then becomes possible to take in the corresponding expressions
the limit in which the initial constant spacetime fields are null, without generating any singularity in the new RS vectors $\vec{R}$ and $\vec{S}$.
Then as a last transformation, apply to these RS vectors a constant but pure imaginary time shift, to generate the final electromagnetic
configuration of fields. In MLE case this series of transformations, starting then from a null static and homogeneous one, produces the
hopfion-Ra\~nada knots. In the present case they produce a doublet of ModMax deformed hopfion-Ra\~nada knots distinguished by $\delta=\pm 1$,
each of which reduces back continuously to the hopfion-Ra\~nada knot when $\gamma\rightarrow 0$.

\subsection{A Static and Homogeneous Field Configuration}
\label{Sect4.1}

Consider the following static and homogeneous $(\vec{E},\vec{B})$ field configuration in the ModMax Lagrangian formulation,
\begin{equation}
\vec{E}=E_0\,\hat{e}_1,\qquad \vec{B}=B_0\,\hat{e}_2,
\end{equation}
such that $\vec{E}$ and $\vec{B}$ are perpendicular from the outset, but not yet necessarily of equal norm,
\begin{equation}
{\cal S}=\frac{1}{2}(E^2_0-B^2_0)\ne 0,\qquad {\cal P}=0.
\end{equation}
Keeping $|E_0|$ and $|B_0|$ of different values is a form of a regularisation in the case of a null field configuration.
Hereafter it turns out it is only the sign of ${\cal S}$ which matters and not its value as such. In a likewise manner
one may choose a static and homogeneous field configuration such that this time ${\cal S}=0$ but ${\cal P}\ne 0$, thus with the two fields of equal norm
but not being exactly perpendicular. In that case as well, the sign of $\vec{E}\cdot\vec{B}$ would prove to be the only parameter
of relevance, and leading to the same results as those presented hereafter once these two fields are taken to be exactly perpendicular.

From the ModMax Lagrangian formulation we have,
\begin{equation}
\vec{D}=\frac{\partial{\cal L}}{\partial\vec{E}}=\cosh\gamma\,\vec{E}\,+\,\sinh\gamma
\frac{(\vec{E}^2-\vec{B}^2)\vec{E} + 2(\vec{E}\cdot\vec{B})\vec{B}}
{\sqrt{(\vec{E}^2-\vec{B}^2)^2+4(\vec{E}\cdot\vec{B})^2}},
\end{equation}
\begin{equation}
\vec{H}=-\frac{\partial{\cal L}}{\partial\vec{B}}=\cosh\gamma\,\vec{B}\,+\,\sinh\gamma
\frac{(\vec{E}^2-\vec{B}^2)\vec{B} - 2(\vec{E}\cdot\vec{B})\vec{E}}
{\sqrt{(\vec{E}^2-\vec{B}^2)^2+4(\vec{E}\cdot\vec{B})^2}}.
\end{equation}
For the field configuration under consideration one then finds
\begin{equation}
\vec{D}=e^{\delta\gamma}\,E_0\,\hat{e}_1,\qquad
\vec{H}=e^{\delta\gamma}\,B_0\hat{e}_2,\qquad {\rm with}\ \ \ \delta={\rm sign}\,(E^2_0-B^2_0)=\pm 1.
\end{equation}
These fields are
such that $\vec{D}\cdot\vec{H}=0$ and $\vec{D}^2-\vec{H}^2=e^{2\delta\gamma}(E^2_0-B^2_0)\ne 0$, while
\begin{equation}
\vec{R}=e^{\delta \gamma} E_0 \hat{e}_1 + i B_0 \hat{e}_2,\qquad
\vec{S}=e^{\delta \gamma} B_0 \hat{e}_2 - i E_0 \hat{e}_1.
\end{equation}
Quite obviously these RS fields $\vec{R}$ and $\vec{S}$ obey the required NLE equations of motion.

\subsection{The Conformal Inversion}
\label{Sect4.2}

Since ModMax equations of motion are covariant under spacetime conformal transformations, a conformal transformation
of the above static and homogeneous field configuration produces yet again a solution to the ModMax equations,
and with a value for the RS vector $\vec{S}$ such that the relation (\ref{eq:S-ModMax}) between $\vec{S}$ and $\vec{R}$ required by the ModMax Hamiltonian
density is indeed met. In particular under a coordinate transformation $x^\mu\rightarrow\tilde{x}^\mu(x^\mu)$ we have for the transformed field strength
$F_{\mu\nu}$,
\begin{equation}
\tilde{F}_{\mu\nu}(\tilde{x})=\frac{\partial x^\rho}{\partial\tilde{x}^\mu}\,\frac{\partial x^\sigma}{\partial\tilde{x}^\nu}\,F_{\rho\sigma}(x).
\end{equation}

Thus consider now specifically the following conformal inversion, with scale factor $\lambda_0>0$ and the notations $\underline{x}^2\equiv x\cdot x=x_\mu x^\mu$,
$\tilde{\underline{x}}^2\equiv\tilde{x}\cdot\tilde{x}=\tilde{x}_\mu \tilde{x}^\mu$,
\begin{equation}
\tilde{x}^\mu=\lambda^2_0\frac{x^\mu}{\underline{x}^2},\qquad
x^\mu=\lambda^2_0\frac{\tilde{x}^\mu}{\tilde{\underline{x}}^2},\qquad
\frac{\partial x^\rho}{\partial\tilde{x}^\mu}=\frac{\lambda^2_0}{(\tilde{\underline{x}}^2)^2}\left(\tilde{\underline{x}}^2\,\delta^\rho_\mu - 2 \tilde{x}_\mu \tilde{x}^\rho\right).
\end{equation}
Using the fact that $F_{0i}=E^i$ and $F_{ij}=-\epsilon^{ijk} B^k$,
a patient calculation establishes the following expressions for the transformed electromagnetic fields. In these expressions the transformed
variables $\tilde{x}^\mu$ are already written simply as $x^\mu=(t,x,y,z)$. One then finds,
\begin{equation}
\vec{E}(x^\mu)=\frac{\lambda^4_0}{(t^2-\vec{x}\,^2)^3}\left(
\begin{array}{c}
E_0(x^2-y^2-z^2-t^2)+B_0(2tz) \\
E_0 (2 xy) \\
E_0(2xz) + B_0(-2tx) \end{array}\right),
\end{equation}
\begin{equation}
\vec{B}(x^\mu)=\frac{\lambda^4_0}{(t^2-\vec{x}\,^2)^3}\left(
\begin{array}{c}
B_0(-2xy) \\
E_0 (-2tz) + B_0(x^2-y^2+z^2+t^2) \\
E_0(2ty) + B_0(-2yz) \end{array}\right).
\end{equation}

A direct calculation then finds for the transformed fields, as one ought to expect given the transformation properties of ${\cal S}$ and ${\cal P}$
under conformal transformations,
\begin{equation}
\vec{E}\cdot\vec{B}=0,\quad
\vec{E}^2-\vec{B}^2=\frac{\lambda^8_0}{(t^2-\vec{x}\,^2)^4}(E^2_0 - B^2_0),\quad
{\rm sign}\,(\vec{E}^2-\vec{B}^2)={\rm sign}\,(E^2_0-B^2_0)=\delta=\pm 1.
\end{equation}
Furthermore from their definitions in the Lagrangian formulation, correspondingly we have for the fields $\vec{D}$ and $\vec{H}$,
\begin{equation}
\vec{D}(x^\mu)=e^{\delta\gamma}\,\vec{E}(x^\mu),\qquad
\vec{H}(x^\mu)=e^{\delta\gamma}\,\vec{B}(x^\mu).
\end{equation}

Since in the remainder of the calculations there is no risk of running into a singularity by setting from hereon $E^2_0-B^2_0=0$ --- the sign $\delta$ being
the sole parameter remaining from the above approach --- let us choose now the following normalisation for $E_0$,
\begin{equation}
E_0 = - \eta\,B_0,\qquad \eta=\pm 1.
\end{equation}
Note that given this choice of normalisation both pairs of fields $(\vec{E},\vec{B})$ and $(\vec{D},\vec{H})$,
whether for the static and homogeneous configuration or for the present one, are null fields.

Finally since the next and final transformation to be effected is the pure imaginary shift in the time coordinate with the use of the Riemann-Silberstein vectors,
let us express the latter two vector quantities for the present two field configurations distinguished by $\delta=\pm 1$,
\begin{equation}
\vec{R}(x^\mu)=\frac{\lambda^4_0}{(t^2-\vec{x}\,^2)^3}B_0\left(
\begin{array}{c}
e^{\delta\gamma}\left[-\eta(x^2-y^2) + \eta(z+\eta t)^2\right] - 2i xy \\
e^{\delta\gamma}\left[-2\eta  xy\right] + i \left[x^2 - y^2 + (z + \eta t)^2 \right] \\
e^{\delta\gamma}\left[-2\eta x(z + \eta t)\right] - 2 i y(z+\eta t) \end{array}\right),
\end{equation}
\begin{equation}
\vec{S}(x^\mu)=\frac{\lambda^4_0}{(t^2-\vec{x}\,^2)^3}B_0\left(
\begin{array}{c}
e^{\delta\gamma}\left[-2xy\right] - i \left[-\eta(x^2-y^2)+\eta(z+\eta t)^2\right] \\
e^{\delta\gamma}\left[x^2-y^2+(z+\eta t)^2\right] + 2i\eta xy \\
e^{\delta\gamma}\left[-2y(z+\eta t)\right] + 2i\eta x(z+\eta t) \end{array}\right).
\end{equation}
Through a direct calculation one may check that indeed,
\begin{equation}
\vec{\nabla}\cdot\vec{R}=0,\qquad
\vec{\nabla}\times\vec{S}=\partial_t\vec{R},
\label{eq:EOM}
\end{equation}
while of course,
\begin{equation}
\vec{R}(x^\mu)=e^{\delta\gamma}\,\vec{E}(x^\mu) + i \vec{B}(x^\mu),\qquad
\vec{S}(x^\mu)=e^{\delta\gamma}\,\vec{B}(x^\mu)-i\vec{E}(x^\mu).
\end{equation}

\subsection{The Pure Imaginary Time Shift and the ModMax deformed HR knots}
\label{sect4.3}

Let us now consider the change of variable\footnote{We recall that the choice of units is such that $c=1$.}
\begin{equation}
t\longrightarrow t- iL,
\end{equation}
where $L$ is some length scale. It proves useful to rescale as well all space-time coordinates as follows,
\begin{equation}
(t,x,y,z)=L\,(T,X,Y,Z).
\end{equation}
For reasons discussed above it should be obvious that if $(\vec{R}(t,x,y,z,),\vec{S}(t,x,y,z))$ obey the equations of motion (\ref{eq:EOM}), so do
the fields $(\vec{R}(t-iL,x,y,z),\vec{S}(t-iL,x,y,z))$. Hence once one separates the real and imaginary parts of the thereby
transformed Riemann-Silberstein vectors one obtains fields for $\vec{E}$, $\vec{B}$, $\vec{D}$ and $\vec{H}$
which again obey the proper equations of motion, inclusive of the necessary relation (\ref{eq:S-ModMax}) between these fields as dictated
by the ModMax Lagrangian and Hamiltonian densities.

When applying the above rescaling it turns out that the factor proportional to $B_0$ setting the amplitude of these four new fields
combines in the form
\begin{equation}
\left(\frac{\lambda_0}{L}\right)^4\,B_0=B.
\end{equation}
The solutions hereafter are thus given in terms of the quantity\footnote{When choosing the same scale factors for both
the conformal inversion and the pure imaginary time shift, $\lambda_0=L$, which would seem just natural since these scale factors
are arbitrary anyway, one has $B=B_0$.} $B$.

In order to express the results of this last transformation, let us introduce the following combinations and notations,
\begin{equation}
A=\frac{1}{2}(1+X^2+Y^2+Z^2-T^2),\quad
(A-iT)^3=Q+iP,\quad
Q=A(A^2-3T^2),\quad
P=T(T^2-3A^2).
\end{equation}
Note that we have $P^2+Q^2=(A^2+T^2)^3$.
It also proves useful to introduce the following collection of four vectors $\vec{H}_\alpha$ ($\alpha=1,2,3,4$), decomposed in the orthonormal
cartesian frame $(\hat{e}_1,\hat{e}_2,\hat{e}_3)$ related to the coordinate system being used,
\begin{equation}
\vec{H}_1:\ \left(
\begin{array}{c}
-XY-e^{\delta\gamma}(Z+\eta T) \\
-\frac{1}{2}\left[1-X^2+Y^2-(Z+\eta T)^2\right] \\
e^{\delta\gamma}X - Y(Z+\eta T)
\end{array}\right),\quad
\vec{H}_2:\ \left(
\begin{array}{c}
\frac{1}{2}\left[1+X^2-Y^2-(Z+\eta T)^2\right] \\
XY - e^{\delta \gamma}(Z+\eta T) \\
e^{\delta\gamma}Y + X(Z+\eta T)
\end{array}\right),
\end{equation}
\begin{equation}
\vec{H}_3:\ \left(
\begin{array}{c}
-e^{\delta\gamma}XY-(Z+\eta T) \\
-\frac{1}{2}e^{\delta\gamma}\left[1-X^2+Y^2-(Z+\eta T)^2\right] \\
X - e^{\delta\gamma}Y(Z+\eta T)
\end{array}\right),\quad
\vec{H}_4:\ \left(
\begin{array}{c}
\frac{1}{2}e^{\delta\gamma}\left[1+X^2-Y^2-(Z+\eta T)^2\right] \\
e^{\delta\gamma}XY - (Z+\eta T) \\
Y + e^{\delta\gamma}X(Z+\eta T)
\end{array}\right).
\end{equation}
All the components of these four vectors are real quantities.
In addition when $\gamma=0$ we have $\vec{H}_3=\vec{H}_1$ and $\vec{H}_4=\vec{H}_2$, while the remaining two vectors
then coincide with those relevant to represent the ordinary hopfion-Ra\~nada knot\cite{Review1}.

A detailed analysis of the above pure imaginary time shift transformation then finds, for the transformed RS vectors,
\begin{equation}
\vec{R}=-\frac{1}{4}iB\frac{1}{(A+iT)^3}\left(\vec{H}_1 + i\eta\vec{H}_4\right) = \vec{D} + i \vec{B},
\end{equation}
\begin{equation}
\vec{S}=-\frac{1}{4}B\frac{1}{(A+iT)^3}\left(\vec{H}_3 + i\eta\vec{H}_2\right)= \vec{H} - i \vec{E}.
\end{equation}
Consequently the electromagnetic fields thereby identified and which obey --- as may be checked explicitly by direct calculation as well --- the correct
equations of motion and relation (\ref{eq:S-ModMax}) are obtained in the following form,
\begin{equation}
\vec{D}=\frac{1}{4}B\frac{1}{(A^2+T^2)^3}\left(P\vec{H}_1 + \eta Q\vec{H}_4\right),\qquad
\vec{E}=\frac{1}{4}B\frac{1}{(A^2+T^2)^3}\left(P\vec{H}_3+\eta Q\vec{H}_2\right),
\end{equation}
\begin{equation}
\vec{B}=\frac{1}{4}B\frac{1}{(A^2+T^2)^3}\left(-Q\vec{H}_1 + \eta P\vec{H}_4\right),\qquad
\vec{H}=\frac{1}{4}B\frac{1}{(A^2+T^2)^3}\left(-Q\vec{H}_3+\eta P\vec{H}_2\right).
\end{equation}
In the limit $\gamma=0$, not only do we have $\vec{D}=\vec{E}$ and $\vec{H}=\vec{B}$ as it should,
but these expressions coincide as well exactly with those which give, in the same parametrisation, the electromagnetic fields
$\vec{E}$ and $\vec{B}$ of the HR knot\cite{Review1,Review2}. However when $\gamma\ne 0$ in fact there correspond two distinct solutions,
distinguished by $\delta = \pm 1$, solving the NLE ModMax equations and reducing precisely to the HR knot in the limit of MLE.

In order to better understand the properties of these two new knot solutions in ModMax theories,
it is necessary to compute all inner products of the vectors $\vec{H}_\alpha$. After some work one then finds,
\begin{eqnarray}
\vec{H}^2_1 &=& \left[A+\eta T(Z+\eta T)\right]^2 + (e^{2\delta\gamma}-1)\left[X^2+(Z+\eta T)^2\right], \nonumber \\
\vec{H}^2_2&=&\left[A+\eta T(Z+\eta T)\right]^2 + (e^{2\delta\gamma}-1)\left[Y^2+(Z+\eta T)^2\right], \nonumber \\
\vec{H}^2_3&=&e^{2\delta\gamma}\left[A+\eta T(Z+\eta T)\right]^2 - (e^{2\delta\gamma}-1)\left[X^2+(Z+\eta T)^2\right], \nonumber \\
\vec{H}^2_4&=&e^{2\delta\gamma}\left[A+\eta T(Z+\eta T)\right]^2 - (e^{2\delta\gamma}-1)\left[Y^2+(Z+\eta T)^2\right], \nonumber \\
\vec{H}_1\cdot\vec{H}_2&=&(e^{2\delta\gamma}-1)\,XY, \\
\vec{H}_1\cdot\vec{H}_3&=&e^{\delta\gamma}\left[A+\eta T(Z+\eta T)\right]^2, \nonumber \\
\vec{H}_1\cdot\vec{H}_4&=&(e^{2\delta\gamma}-1)(Z+\eta T)\left[A+\eta T(Z+\eta T) -1\right], \nonumber \\
\vec{H}_2\cdot\vec{H}_3&=&-(e^{2\delta\gamma}-1)(Z+\eta T)\left[A+\eta T(Z+\eta T) -1\right], \nonumber \\
\vec{H}_2\cdot\vec{H}_4&=&e^{\delta\gamma}\left[A+\eta T(Z+\eta T)\right]^2, \nonumber \\
\vec{H}_3\cdot\vec{H}_4&=&-(e^{2\delta\gamma}-1)\,XY. \nonumber
\end{eqnarray}
Note that in the limit when $\gamma=0$, all these results are consistent with the fact that $\vec{H}_3$ and $\vec{H}_4$ reduce to $\vec{H}_1$ and
$\vec{H}_2$, respectively, and furthermore that $\vec{H}_1$ and $\vec{H}_2$ are perpendicular and of same norm as is the case for the HR knot.

Given these solutions parametrised by the magnetic field strength scale $B$, the length scale factor $L$, and $\delta=\pm 1$, one may then compute,
\begin{equation}
\frac{1}{2}\left(\vec{E}^2-\vec{B}^2\right)=-\frac{1}{2}(e^{2\delta\gamma}-1)\left(\frac{B}{4}\right)^2\frac{X^2-Y^2}{(A^2+T^2)^3}=
-\frac{1}{2}\left(\vec{D}^2-\vec{H}^2\right),
\end{equation}
as well as,
\begin{equation}
\vec{E}\cdot\vec{B}=-\eta(e^{2\delta\gamma}-1)\left(\frac{B}{4}\right)^2\frac{XY}{(A^2+T^2)^3}=-\vec{H}\cdot\vec{D}.
\end{equation}
Therefore when $\gamma\ne 0$ these solutions do not define null fields for either pair $(\vec{E},\vec{B})$ or $(\vec{D},\vec{H})$.
Yet when $\gamma=0$ indeed one recovers the correct null fields of the HR knot. In particular this means that as a matter of fact,
both these solutions for $\delta=\pm 1$ are well defined without any singularity for both the Lagrangian and Hamiltonian formulations of ModMax theories.

Incidentally one may also compute explicitly that
\begin{equation}
\vec{D}\cdot\vec{E}=\left(\frac{B}{4}\right)^2e^{\delta\gamma}\frac{\left[A+\eta T(Z+\eta T)\right]^2}{(A^2+T^2)^3}=\vec{H}\cdot\vec{B},
\end{equation}
as well as,
\begin{equation}
\vec{R}\cdot\vec{S}=0.
\end{equation}
In particular given the identity
\begin{equation}
\vec{R}\cdot\vec{S}=\left(\vec{D}\cdot\vec{H}+\vec{E}\cdot\vec{B}\right) -i\left(\vec{D}\cdot\vec{E}-\vec{H}\cdot\vec{B}\right),
\end{equation}
all these results for the different scalar products of electric and magnetic fields are indeed consistent.

\subsection{Relativistic Kinematics and Topological Properties}
\label{Sect4.4}

The energy and momentum kinematics of these solutions may also be evaluated, given the definitions,
\begin{equation}
E=\int_{(\infty)}d^3\vec{x}\,{\cal H}_\gamma,\quad
\vec{P}=\int_{(\infty)}d^3\vec{x}\,\vec{D}\times\vec{B},\quad
\vec{D}\times\vec{B}=\eta\left(\frac{B}{4}\right)^2\frac{1}{(A^2+T^2)^3}\,\vec{H}_1\times\vec{H}_4,
\end{equation}
and leading to the final values, after spatial integration,
\begin{equation}
E=\frac{1}{8}\pi^2\,L^3\,e^{\delta\gamma}\,B^2\,\left(1+\frac{1}{2}\sinh^2\gamma\right),\qquad
\vec{P}=\frac{1}{16}\eta\,\pi^2\, L^3\,e^{\delta\gamma}\,B^2\,\hat{e}_3.
\end{equation}
In particular the average velocity and relativistic invariant mass of these ModMax deformed HR knots are,
\begin{equation}
\vec{\beta}=\frac{\vec{P}}{E}=\frac{1}{2}\eta\,\frac{1}{1+\frac{1}{2}\sinh^2\gamma}\,\hat{e}_3  ,\qquad
M=\frac{1}{16}\pi^2\,L^3\,e^{\delta\gamma}\,B^2\,\sqrt{3+(4+\sinh^2\gamma)\sinh^2\gamma}.
\end{equation}

Since the two ModMax deformed HR knots are continuous deformations in the parameter $\gamma$ of the ordinary HR knot,
the topologically nontrivial structure and properties of the closed field lines of the latter are being preserved in the former,
thus sharing the same linking and knotting properties as measured by the electric and magnetic helicities $h_e$, $h_m$ and $h_{em}=h_{me}$
introduced in Sect.\ref{Sect2.2} in terms of the vector potentials $\vec{C}$, $\vec{A}$ and $\vec{\Phi}$ and their derived
fields $\vec{D}$, $\vec{B}$ and $\vec{R}$. Not only by construction as it should, but as may be checked by direct calculation as well,
one has $\vec{\nabla}\cdot\vec{R}=0$ and $\vec{\nabla}\times\vec{S}=\partial_t\vec{R}$, hence in particular
$\vec{\nabla}\cdot\vec{D}=0=\vec{\nabla}\cdot\vec{B}$ which implies that $\vec{D}$ and $\vec{B}$ field lines indeed close
onto themselves. However by direct calculation it may also be checked that for the constructed knot solutions\footnote{One may also compute
$\vec{\nabla}\times\vec{R}$, but the lengthy result is not illuminating.},
\begin{equation}
\vec{\nabla}\cdot\vec{S}=0,\qquad
\vec{\nabla}\cdot\vec{E}=0,\qquad
\vec{\nabla}\cdot\vec{H}=0,
\end{equation}
so that $\vec{E}$ and $\vec{H}$ field lines close onto themselves as well (as should indeed be expected, since by pairs all these field lines
coalesce continuously into those of the ordinary HR null knot when $\gamma=0$ for each of the values $\delta=\pm 1$).

The helicities (\ref{eq:he}), (\ref{eq:hm}) and (\ref{eq:hem}) of the $\vec{D}$ and $\vec{B}$ field lines may be computed in terms of their vector potentials
$\vec{C}$ and $\vec{A}$. In the case of the ordinary HR knot the helicities $h_e$ and $h_m$ are nonvanishing and equal, while $h_{em}=0=h_{me}$,
and are all three conserved under time evolution. Fortunately the necessary vector potentials may readily be identified as well for the knot solutions constructed above,
in terms of an ordinary Bateman representation of these same configurations. Indeed the generalised Bateman potentials may for instance be chosen
in the following form,
\begin{equation}
\alpha(x^\mu)=\frac{L}{2(A+iT)}(Z+i\eta(A-1)),\quad
\beta(x^\mu)=i\,B\,\frac{L}{2(A+iT)}(X+i\eta e^{\delta\gamma} Y),\quad
\sigma^\mu(x^\mu)=0,
\label{eq:potentials-x}
\end{equation}
since, as may be checked, the corresponding vector $\vec{R}=\vec{\nabla}\alpha\times\vec{\nabla}\beta$
then coincides exactly with the one obtained above, while as well $\vec{S}=\partial_t\alpha\vec{\nabla}\beta-\partial_t\beta\vec{\nabla}\alpha$
since we have indeed $\vec{R}\cdot\vec{S}=0$ which requires that $\sigma^\mu=0$.

The identification of these Bateman potentials may be achieved as follows within the ModMax Hamiltonian formulation.
Consider the ordinary Bateman potentials (\ref{eq:Bateman-Constant}) constructed in Sect.\ref{Sect3.1} for static and homogeneous $\vec{D}$ and $\vec{B}$ fields
which are perpendicular, and such that $\sigma^\mu=0$, namely
\begin{equation}
\alpha(u^\mu)=u^3-\eta u^0 -\frac{1}{2}i\eta L,\qquad
\beta(u^\mu)=iB\left(u^1+i\eta e^{\delta\gamma} u^2\right),\qquad \sigma^\mu(u^\mu)=0,
\label{eq:potentials-u}
\end{equation}
where spacetime coordinates $u^\mu$ are used here rather than $x^\mu$, and where specific values for the constants $\alpha_c$ and $\beta_c$
have been chosen for convenience hereafter. Apply then the following change of variables defining a special spacetime conformal transformation
of pure imaginary 4-vector $b^\mu$,
\begin{equation}
u^\mu=\frac{x^\mu + b^\mu\, \underline{x}^2}{1+2\,b\cdot x + \underline{b}^2\,\underline{x}^2},\qquad
b^\mu=\frac{i}{L}(1,0,0,0).
\end{equation}
As is well known this transformation results from the composition of a conformal inversion, followed by a spacetime translation by the 4-vector $b^\mu$,
and then again the inverse of the first conformal inversion. Given the specific chosen values for $\alpha_c$ and $\beta_c$, and upon this special
conformal transformation, the Bateman potentials (\ref{eq:potentials-u}) are transformed into those in (\ref{eq:potentials-x})
for the ModMax knots constructed here.

Furthermore since we may write, for instance,
\begin{equation}
\vec{R}=\vec{\nabla}\alpha\times\vec{\nabla}\beta=\vec{\nabla}\times(\lambda\alpha\vec{\nabla}\beta-(1-\lambda)\beta\vec{\nabla}\alpha),\qquad
\lambda\in\mathbb{C},
\end{equation}
a possible choice of vector potentials is,
\begin{equation}
\vec{C}+i\vec{A}=\lambda\alpha\vec{\nabla}\beta-(1-\lambda)\beta\vec{\nabla}\alpha.
\end{equation}
On the other hand note that one has,
\begin{equation}
(\vec{C}+i\vec{A})\cdot(\vec{D}+i\vec{B})=(\vec{C}\cdot\vec{D}-\vec{A}\cdot\vec{B})+i(\vec{C}\cdot\vec{B}+\vec{A}\cdot\vec{D}),
\end{equation}
while the same quantity is given by,
\begin{equation}
(\vec{C}+i\vec{A})\cdot(\vec{D}+i\vec{B})=(\lambda\alpha\vec{\nabla}\beta-(1-\lambda)\beta\vec{\nabla}\alpha)\cdot(\vec{\nabla}\alpha\times\vec{\nabla}\beta)=0.
\end{equation}
Therefore $\vec{C}\cdot\vec{D}=\vec{A}\cdot\vec{B}$ and $\vec{C}\cdot\vec{B}=-\vec{A}\cdot\vec{D}$. In particular the first of these identities
implies that the electric and magnetic helicities of the constructed knots are equal, $h_e=h_m$, while their total charge under duality transformations
takes the value $Q_{\rm dual}=2h_e$.

Regarding the time dependencies of these helicities it was established in (\ref{eq:dt-h}) that these are governed by
\begin{eqnarray}
\frac{d}{dt}h_e &=& \int_{(\infty)}d^3\vec{x}\,\vec{H}\cdot\vec{D},\qquad
\frac{d}{dt}h_m=-\int_{(\infty)}d^3\vec{x}\,\vec{E}\cdot\vec{B}, \nonumber \\
\frac{d}{dt}h_{em} &=& \frac{d}{dt}h_{me} = \frac{1}{2}\int_{(\infty)}d^3\vec{x}\left(\vec{H}\cdot\vec{B}-\vec{E}\cdot\vec{D}\right).
\end{eqnarray}
In the ordinary HR case the fields being null the densities to be integrated vanish identically. When $\gamma\ne 0$ these densities
no longer vanish identically. However in terms of the explicit expressions for $\vec{H}\cdot\vec{D}=-\vec{E}\cdot\vec{B}$ and
$\vec{D}\cdot\vec{E}-\vec{H}\cdot\vec{B}=0$ given above as they apply to these knot solutions,
the time variations of the three helicities vanish nevertheless, and this upon integration in the case of $h_e=h_m$ since the corresponding
densities are odd in $X$ and $Y$ separately. Hence indeed all three topological helicities characteristic of the constructed solutions are conserved
under time evolution.

That these helicities are nonvanishing and time independent may be confirmed through a direct evaluation of the relevant integrals
for any value of $T$, leading to
\begin{equation}
\frac{1}{2}Q_{\rm dual}=h_e=h_m=\frac{1}{32}\pi^2\,L^3\,e^{\delta\gamma}\,B^2,\qquad
h_{em}=0=h_{me}.
\end{equation}
Indeed for the ordinary HR knot the mixed helicities $h_{em}=h_{me}$ vanish identically as well, while the values for $h_e=h_m$ then coincide with the value
above when $\gamma=0$\cite{Review2}. The nonvanishing values for $h_e$ and $h_m$ confirm the nontrivial topological character of the two ModMax knot configurations
constructed here.

The value of the duality charge scales with the combination $L^3 B^2 e^{\delta\gamma}$ of the parameters involved in these knots, as do other
conserved physical quantities of interest. Since the duality charge $Q_{\rm dual}$ is a Lorentz scalar, let us normalise it to some other Lorentz scalar
characteristic of the solution, namely its relativistic invariant mass. One then finds,
\begin{equation}
\frac{1}{2}\frac{Q_{\rm dual}}{M}=\frac{h_e}{M}=\frac{h_m}{M}=\frac{1}{2}\frac{1}{\sqrt{3+(4+\sinh^2\gamma)\sinh^2\gamma}}.
\end{equation}
In the limit $\gamma=0$ this result coincides with that obtained for the hopfion-Ra\~nada knot.

Finally in order to visualize the closed and linked field lines of the explicit deformed HR solutions obtained above, if only abstractly --- figures are not included here ---
but by comparing to graphs displaying the closed and linked field lines of the ordinary hopfion-Ra\~nada knot with $\gamma=0$
to be found in Refs.\cite{Review1,Review2},
let us consider these field configurations specifically at $T=0$. One then finds\footnote{Note how a $\pi/2$ rotation in the $XY$ plane around
the $Z$ axis which effects the permutations $(X,Y)\rightarrow(Y,-X)$, combined with a $\pi/2$ duality transformation which exchanges
the fields $(\vec{D},\vec{B})\rightarrow(\vec{B},-\vec{D})$ and $(\vec{E},\vec{H})\rightarrow(\vec{H},-\vec{E})$, and with an orientation dependent on
$\eta=\pm 1$, leaves this configuration of fields invariant.}
\begin{eqnarray}
\vec{D}(T=0,X,Y,Z) &=& 2\eta\,B\,\frac{1}{(1+X^2+Y^2+Z^2)^3}
\left(\begin{array}{c}
\frac{1}{2}e^{\delta\gamma}(1+X^2-Y^2-Z^2) \\
e^{\delta\gamma} XY  - Z \\
Y + e^{\delta\gamma} XZ
\end{array}\right) , \nonumber \\
\vec{E}(T=0,X,Y,Z) &=& 2\eta\,B\,\frac{1}{(1+X^2+Y^2+Z^2)^3}
\left(\begin{array}{c}
\frac{1}{2}(1+X^2-Y^2-Z^2) \\
XY - e^{\delta\gamma} Z \\
e^{\delta\gamma} Y + XZ
\end{array}\right) , \nonumber \\
\vec{B}(T=0,X,Y,Z) &=& -2\,B\,\frac{1}{(1+X^2+Y^2+Z^2)^3}
\left(\begin{array}{c}
-XY - e^{\delta\gamma} Z \\
-\frac{1}{2}(1-X^2+Y^2-Z^2) \\
e^{\delta\gamma}X - YZ
\end{array}\right) , \nonumber \\
\vec{H}(T=0,X,Y,Z) &=& -2\,B\,\frac{1}{(1+X^2+Y^2+Z^2)^3}
\left(\begin{array}{c}
-e^{\delta\gamma} XY - Z \\
-\frac{1}{2}e^{\delta\gamma}(1-X^2+Y^2-Z^2) \\
X - e^{\delta\gamma} YZ
\end{array}\right),
\end{eqnarray}
and in particular then in the plane $Z=0$,
\begin{eqnarray}
\vec{D}(T=0,X,Y,Z=0) &=& 2\eta\,B\,\frac{1}{(1+X^2+Y^2)^3}
\left(\begin{array}{c}
\frac{1}{2}e^{\delta\gamma}(1+X^2-Y^2) \\
e^{\delta\gamma} XY \\
Y
\end{array}\right) , \nonumber \\
\vec{E}(T=0,X,Y,Z=0) &=& 2\eta\,B\,\frac{1}{(1+X^2+Y^2)^3}
\left(\begin{array}{c}
\frac{1}{2}(1+X^2-Y^2) \\
XY \\
e^{\delta\gamma} Y
\end{array}\right) , \nonumber \\
\vec{B}(T=0,X,Y,Z=0) &=& -2\,B\,\frac{1}{(1+X^2+Y^2)^3}
\left(\begin{array}{c}
-XY \\
-\frac{1}{2}(1-X^2+Y^2) \\
e^{\delta\gamma}X
\end{array}\right) , \nonumber \\
\vec{H}(T=0,X,Y,Z=0) &=& -2\,B\,\frac{1}{(1+X^2+Y^2)^3}
\left(\begin{array}{c}
-e^{\delta\gamma} XY \\
-\frac{1}{2}e^{\delta\gamma}(1-X^2+Y^2) \\
X
\end{array}\right).
\end{eqnarray}
Whatever the value for $\delta=\pm 1$, and depending on the value for the ModMax deformation parameter $\gamma\ge 0$ contributing through the sole
quantity $e^{\delta\gamma}$ involved in these expressions, these field configurations display the following behaviour when $\gamma$ is turned on.
The closed and linked $\vec{E}$ and $\vec{B}$ field lines of the ordinary HR solution at $\gamma=0$ are then each split, when $\gamma\ne 0$, into
pairs of closed and linked $\vec{D}$ and $\vec{E}$, on the one hand, and $\vec{B}$ and $\vec{H}$, on the other hand, field lines, of which some
of the 3-vector components have some of their contributions rescaled by the factor $e^{\delta\gamma}$ and this in a manner dependent both
on the considered point $(X,Y,Z)$ in space and on which field is being evaluated.
In other words, as compared to their configuration when $\gamma=0$, when $\gamma$ is turned on all field lines are
tilted and twisted in this or that direction in space in a manner dependent on the point in space which that field line is crossing ---
yet by maintaining however, the topologically nontrivial loop structure of all these closed field lines which, for $\gamma=0$, wind around
embedded circular torii all sharing the $Z$ axis as a common rotational symmetry axis\cite{Review1,Review2},
these torii thus being deformed by twisting and squeezing when $\gamma\ne 0$ while preserving their embedded topology to still fill up all of space.
In particular what would be a $\pi/2$ circular rotation symmetry around the $Z$ axis when $\gamma=0$, is deformed
when $\gamma\ne 0$ into a shape that reminds one of an elliptic-type deformation along certain principal axes (given the above expressions at $T=0$
this is easiest to visualize in the $Z=0$ plane). This feature of these ModMax deformed HR knots is the analogue of what happens
for transverse circularly polarised travelling plane waves for $\gamma=0$ being deformed into elliptically polarised ones when $\gamma\ne 0$,
as pointed out\cite{Town} in Sect.\ref{Sect3.2}.

\section{Conclusions}
\label{Sect5}

Source-free ModMax theories\cite{Town} of nonlinear electrodynamics (NLE) in the four dimensional Minkowski spacetime vacuum
are the only possible continuous deformations of source-free Max\-well linear electrodynamics (MLE) in the same vacuum,
which preserve all the same continuous Poincar\'e and conformal spacetime symmetries and duality invariances of MLE.
These ModMax theories are labelled by a single real and positive parameter $\gamma\ge 0$, such that ordinary Maxwell theory
is recovered for $\gamma=0$. Null field configurations play a central role for MLE. They include not only monochromatic travelling plane waves,
for instance, but null electromagnetic knots as well with their topologically nontrivial structures, among which hopfion-Ra\~nada knots constitute
a class of their own. {\sl A priori} any such null configuration remains, without deformation, an exact solution for any Poincar\'e invariant NLE theory,
provided however that spacetime conformal invariance is no longer in place. By lack of an intrinsic physical scale, in the case of a spacetime Poincar\'e
and conformal invariant NLE, null field configurations are ill-defined within the Lagrangian formulation, because of a lack
of analyticity in that case of the Lagrangian density as a function of the electromagnetic Lorentz invariants. Nonetheless because of their topologically
nontrivial structure null knots are expected to remain robust against whatever continuous nonlinear deformations of MLE, which they are indeed 
without deformation when spacetime conformal invariance is not enforced.

After a review of different possible methods to tackle the nonlinearities of the equations of motion of NLE and based on generalisations of the
well established Riemann-Silberstein and Bateman approaches in the case of MLE, the fate within ModMax theories specifically
of the ordinary hopfion-Ra\~nada (HR) knots of MLE has been addressed in this work. It has been established that given any HR knot in MLE characterised
by a magnetic field scale $B$ for its magnetic field component and a length scale $L$ for its spatial extent in physical space,
there correspond to it in ModMax theories two physically distinct deformed HR knots with their preserved nontrivial topological structures,
distinguished by a single discrete parameter $\delta=\pm 1$, and with as continuous deformation variable the quantity $e^{\delta\gamma}$.
While in the limit $\gamma=0$, these two ModMax deformed HR knots coalesce back in a continuous fashion
to the ordinary HR knot of MLE with the same parameters $B$ and $L$. Furthermore and with an interesting twist to what could have been expected,
these ordinary null HR knots remain robust even for such conformally invariant deformations of MLE, by self-adapting their initially null local structure
of their topologically
nontrivial properties. The singular character of the ordinary null HR knots for the ModMax Lagrangian formulation is avoided by the ModMax deformed
HR knots simply by no longer being null electromagnetic field configurations, while maintaining nonetheless in a continuous fashion their topological structure
under deformation in the variable $e^{\delta\gamma}$. Answers to all the questions raised in the Introduction have thus been clarified.

Other issues of interest remain to be understood further, however. The original hopfion-Ra\~nada knots may also be constructed in terms of a double
Hopf fibration of the 3-sphere\cite{HR1,HR2,Review1,Review2}. By continuity the new ModMax deformed HR knots constructed in this work ought also to
correspond to a continuous deformation of that double Hopf fibration of the 3-sphere. The possibility of such a continuous deformation of Hopf fibrations
in relation to ModMax theories deserves to be explored further.

As pointed out in Sect.\ref{Sect4.4}, the ModMax deformed HR knots possess a representation in terms of the ordinary Bateman construction
based on two Bateman 4-scalar potentials $(\alpha,\beta)$, and a vanishing Bateman 4-vector potential $\sigma^\mu$. Under such circumstances and as discussed
above, any bi-holomorphic transformation of the scalar potentials, $(f(\alpha,\beta),g(\alpha,\beta))$, then defines yet another solution to the ModMax dynamics.
Within the context of MLE and starting from the Bateman scalar potentials for the HR knot, choosing $f(\alpha,\beta)=\alpha^p$ and $g(\alpha,\beta)=\beta^q$
with $(p,q)$ any pair of co-prime positive natural numbers produces\cite{BB2,Review1,Review2} an entire double infinite discrete series of additional
null electromagnetic knots with their own characteristic nontrivial topological properties and helicities. Starting from the doublet of ModMax deformed
HR knots constructed in this paper and their Bateman scalar potentials $(\alpha,\beta)$,
an extensive study of all these ModMax deformed $(p,q)$ knots and their properties could also be developed.

The existence and properties of topologically nontrivial electromagnetic field configurations within generalisations of ordinary linear Maxwell theory
thus remains certainly a topic of fascinating interest, to be pursued further, with its possible implications for the physics of nonlinear electrodynamics
phenomena within different contexts. Hopfion solitons find their mathematical and physical relevance in a growing number of different fields beyond
linear electromagnetism, ranging among others from condensed quantum matter physics to magnetohydrodynamics (see for instance Ref.\cite{Hopfion}).
Furthermore, nonlinear electrodynamics is witnessing an increased interest as possibly playing a relevant role in the dynamics of cosmological expansion
(see for example  Refs.\cite{Novello,Kruglov1,Kruglov2}). ModMax theories lend themselves to some very sensitive experimental tests
of vacuum birefringence\cite{Denisov,Town}, while quantum corrections are known to lead to nonlinear corrections to Maxwell's equations as in the
Heisenberg-Euler Lagrangian, albeit not necessarily conformally or duality invariant
ones. When coupled to a curved spacetime metric with its intrinsic nonlinear dynamics, nonlinear back reaction corrections to ordinary Maxwell
dynamics ought to be expected. Having theoretical models to help explore beyond the boundaries of present day fundamental theories always
enables to better test and understand the latter to their limits. The robustness to nonlinear deformations of hopfion-Ra\~nada knots and other topologically
nontrivial electromagnetic configurations, even known as exact analytic solutions to some classes of nonlinear extensions of ordinary Maxwell theory,
makes them ideal beacons with which to project our gaze deeper into today's unknown territories at the frontiers.

\section*{Acknowledgements}

The work of CD and JG is supported in part by the Institut Interuniversitaire des Sciences
Nucl\'eaires (IISN, Belgium).

\end{document}